\documentclass[twocolumn]{aastex63}

\usepackage{amsmath}
\usepackage{amsbsy}
\let\oldAA\AA
\newcommand\angst{\text{\normalfont\oldAA}}
\newcommand\nh{\ifmmode{n_{\tiny \mbox{H}}}\else{$n_{\tiny \mbox{H}}$}\fi}
\newcommand\ngr{\ifmmode{n_{\tiny \mbox{gr}}}\else{$n_{\tiny \mbox{gr}}$}\fi}
\newcommand\jvsp{\ifmmode{j_{\tiny \nu,sp}}\else{$j_{\tiny \nu, sp}$}\fi}
\newcommand\jvff{\ifmmode{j_{\tiny \nu,ff}}\else{$j_{\tiny \nu,ff}$}\fi}
\newcommand\jvbb{\ifmmode{j_{\tiny \nu,bb}}\else{$j_{\tiny \nu,bd}$}\fi}
\newcommand\amax{\if{a_{\tiny \mbox{max}}}\else{$a_{\tiny \mbox{max}}$}\fi}
\newcommand\amin{\if{a_{\tiny \mbox{min}}}\else{$a_{\tiny \mbox{min}}$}\fi}
\newcommand\cmvol{\ifmmode{\mbox{cm}^{-3}}\else{$\mbox{cm}^{-3}$}\fi}
\newcommand\HII{$\textrm{H} \scriptstyle\mathrm{II}$}

\received{September 22, 2021}
\revised{June 29, 2022}
\accepted{July 25, 2022}
\submitjournal{ApJ}

\shorttitle{Radio-\textit{mm} SED of SF region with AME}
\shortauthors{Yoon}
\graphicspath{{./}{figures/}}

\begin{document}

\title{Spinning Nanoparticles Impacted by C-shock: Implications for Radio-millimeter Emission from Star-forming Regions}

\correspondingauthor{Ilsang Yoon}
\email{iyoon@nrao.edu}

\author[0000-0001-9163-0064]{Ilsang Yoon}
\affiliation{National Radio Astronomy Observatory, 520 Edgemont Road, Charlottesville, VA 22903, USA}











\begin{abstract}
We investigate the impact of anomalous microwave emission (AME) on the radio-millimeter spectral energy distribution for three typical interstellar medium (ISM) conditions surrounding star-forming regions -- cold neutral medium, warm neutral medium, and photodissociation region -- by comparing the emissivities of three major contributors: free-free, thermal dust emission, and AME. In particular, for spinning nanoparticles (i.e., potential carriers of AME), we consider a known grain destruction mechanism due to a centrifugal force from spin-up processes caused by collisions between dust grains and supersonic neutral streams in a magnetized shock (C-shock). We demonstrate that, if the ISM in a magnetic field is impacted by a C-shock developed by a supernova explosion in the early phase of massive star-formation ($\lesssim 10$ Myr), AME can be significantly or almost entirely suppressed relative to free-free and thermal dust continuum emission if the grain tensile strength is small enough. This study may shed light on explaining the rare observations of AME from extragalactic star-forming regions preferentially observed from massive star clusters and suggest a scenario of ``the rise and fall of AME'' in accordance with the temporal evolution of star-forming regions. 
\end{abstract}

\keywords{submillimeter: ISM -- radio continuum: ISM -- galaxies: ISM -- ISM:dust -- radiation mechanisms: non-thermal -- radiation mechanisms: thermal}

\section{Introduction} \label{sec:intro}
Thermal free-free emission is a good tracer of star formation and widely used to estimate the star formation rate (SFR) of galaxies using radio continuum measurements at 10-33 GHz where the free-free emission is dominant \citep[][]{murphy_etal_2011}. However, anomalous microwave emission (AME) explained by dipole radiation from spinning nanoparticles \citep{draine_and_lazarian_1998} is also bright at this frequency range, and the peak frequency can be as high as 100 GHz \citep[][]{spdust2009}, which implies that (1) AME can be a significant contamination if the SFR is measured by thermal free-free radio continuum emission from a single frequency \citep[e.g.,][]{murphy_etal_2011} and (2) AME can affect a measurement of molecular gas mass by a submillimeter (e.g., 850 $\mu$m or 350 GHz) flux density \citep[e.g.,][]{scoville_etal_2016}.    

Therefore, the observational study of AME in Galactic and extragalactic star-forming regions is important to understand the impact of AME on the radio-millimeter continuum spectral energy distribution (SED). The current estimate of the fraction of AME to the total emission at a frequency near $\approx 30$ GHz in our Galaxy is as large as 50\% \citep{dickinson_etal_2018}. However, in other galaxies, AME does not seem to be strong, as shown by the observations of two galaxies, NGC 6946 \citep{murphy_etal_2010,scaife_etal_2010b,hensley_etal_2015} and NGC 4725B \citep{murphy_etal_2018} although the integrated flux over entire M31 galaxy shows significant enhancement of AME \citep{battistelli_etal_2019}.

The extragalactic AME detection is reported for only two galaxies (NGC 6946 and NGC 4725B) with spatially resolved emission and one galaxy (M31) with spatially integrated emission. It is not clear whether this discrepancy is due to a different interstellar medium (ISM) environment associated with the AME region in our Galaxy and other galaxies or just a bias in the AME estimate in our Galaxy based on the observation of solar neighbors or a combination of both \citep{dickinson_etal_2018}. If the observation of AME in our Milky Way is biased, and AME is indeed weak and rare in star-forming regions in galaxies in general, it is interesting to understand the possible reasons for the weakness and rareness of AME in extragalactic star-forming regions.  

In general, a star-forming region is complex and consists of a mixture of different phases of ISM from the fully ionized phase (\HII\ region) to the cold phase of gas and dust shielding strong UV radiation, as illustrated in Figure~\ref{fig:sfregion}. As illustrated by \cite{spdust2009} using models of spinning dust emission, the types of ISM from which one can expect both thermal free-free emission and significant AME are cold neutral medium (CNM), warm neutral medium (WNM), and photodissociation region (PDR) that are usually associated with nearby \HII\ regions (Figure~\ref{fig:sfregion}). 

Although our current understanding of a connection between the physical ISM conditions of star-forming regions and the AME detection is still incomplete and evolving, the emerging impression is that the lower-density star-forming regions with a high interstellar radiation field (ISRF; like WNM) provide the most promising environment for detecting AME, while a very high ISRF may destroy the grain population \citep[see ][for a review]{scaife_2013}. 

Recent observations of the resolved star-forming clouds suggest that AME is correlated with polycyclic aromatic hydrocarbon (PAH) emission associated with the PDR \citep[][]{bell_etal_2019,arcetord_etal_2020,casassus_etal_2021}, which is consistent with the previous observations suggesting a correlation between AME and PAH tracers \citep{scaife_etal_2010a,ysard_etal_2010,tibbs_etal_2011,battistelli_etal_2015}.

The AME may also be detected in \HII\ regions \citep[e.g., RCW 175 by][]{tibbs_etal_2012,battistelli_etal_2015} if the dust is present in the interior of the \HII\ region \citep{paladini_etal_2012,watson_etal_2008} or in the \HII\ bubble \citep{anderson_etal_2012,flagey_etal_2011}. However, in general, it is not likely that AME in the \HII\ regions dominates the free-free emission because the abundance of dust particles (e.g., PAH) in \HII\ regions is suppressed due to strong UV radiation \citep[e.g.,][]{peeters_etal_2004,binder_and_povich_2018,povich_etal_2007} and supernova shocks \citep[e.g.,][]{jones_etal_1996}, and, even in high-density \HII\ regions with increased dust abundance \citep[e.g.,][]{draine_2011}, free-free emission is likely to be dominant due to the free-free emissivity scaled as $n^2_e$.   

In the current working models of AME \citep[e.g.,][]{draine_and_lazarian_1998,spdust2009}, there are several model parameters -- distribution of the grain size, charge, dipole moment, strength of the ambient ISRF, gas temperature, and hydrogen density \citep[e.g.,][]{spdust2009,hensley_and_draine_2017} -- that can change the shape and intensity of the AME emissivity function such that AME can be suppressed. However, the emissivity of AME from CNM, WNM, and PDR is still significantly larger than the free-free emissivity in the frequency of 10-100 GHz for typical parameter values of the ISM and dust \citep{spdust2009}. Furthermore, the recently proposed dust destruction process by radiative torque in a strong radiation field suggests that it efficiently breaks large grains \citep{hoang_etal_rat_2019} and increases the small-grain population, which implies that AME can be even stronger in those star-forming regions with strong UV radiation from OB star associations (see Section ~\ref{sec:riseame}).

Another possible (and probably more plausible) way of suppressing AME from those star-forming regions (CNM, WNM, and PDR) is to reduce the abundance of small nanoparticles by destroying them, which has not been discussed much. Recently, \cite{hoang_etal_2019} suggested a new disruption mechanism of very small grains (nanoparticles with size $\sim 10^{-9}$m) by increased centrifugal force due to suprathermal rotation by stochastic mechanical torque (originally proposed by \cite{gold_1952}) from a magnetized shock (C-shock). The magnetic field is ubiquitous, and, for the ISM with a lower ionization fraction, the supersonic drift of neutral particles in a C-type shock \citep{draine_2011book} can impact small nanoparticles bound by the magentic field and increase their angular velocity \citep{hoang_etal_2019}. Indeed, relative to other grain destruction mechanisms (thermal sputtering, nonthermal sputtering, and grain--grain collision), this rotational disruption appears to be the fastest mechanism to destroy nanoparticles in C-shocks, as suggested by the comparison of the time scale for each process \citep{hoang_etal_2019}.

In this work, we compute the emissivity of AME by incorporating the disruption of nanoparticles impacted by C-shocks and investigate the emissivity of the composite of AME, thermal free-free, and dust continuum
emission in the radio-millimeter wavelength (10--100 GHz) from the ISM associated with star-forming regions, to characterize how much AME contributes to the radio-millimeter SED.

In Section~\ref{sec:ame}, we introduce a brief overview of the emission mechanism of spinning nanoparticles and the characteristic properties of the ISM associated with star-forming regions. In Section~\ref{sec:dynamics}, the properties of C-shocks and the rotational dynamics of nanoparticles in the region affected by C-shocks are discussed. In Section~\ref{sec:destroy}, we discuss the suprathermal angular rotation velocity of spinning nanoparticles and their critical angular rotation speed to resist centrifugal force, as well as the implications of assumed dust size distribution impacting the AME emissivity. In Section~\ref{sec:emission}, we show the impact of C-shocks on the radio-millimeter emissivity by considering grain destruction mechanisms discussed in Section~\ref{sec:destroy}. In Section~\ref{sec:discuss}, we propose a possible scenario of the rise and fall of AME and discuss the relevant issues related to the current work. Finally, we summarize our work in Section~\ref{sec:summary}.  

\section{AME from Star-forming Regions} \label{sec:ame}
In this section, we provide a brief overview of the grain rotational dynamics. More details are found in \cite{draine_and_lazarian_1998} and \cite{spdust2009}.

\subsection{Basics of Dipole Radiation from Spinning Nanoparticles}\label{sec:ame_instro}
The radiation power at the frequency $\nu=\omega/2\pi$ from spinning particles with an angular velocity $\boldsymbol{\omega}$ of the electric dipole moment $\boldsymbol{\mu}$, with a component $\boldsymbol{\mu}_{\perp}$ perpendicular to $\boldsymbol{\omega}$, is
\begin{equation}
P = \frac{2}{3}\frac{\boldsymbol{\mu}^2_\perp \boldsymbol{\omega}^4}{c^3}.
\end{equation} and the emissivity of the spinning particles per H atom in erg s$^{-1}$ sr$^{-1}$ Hz$^{-1}$ (H atom)$^{-1}$ is 
\begin{equation}\label{eq:emissi}
\frac{\jvsp}{\nh} = \frac{1}{4\pi} \int^{a_{\tiny \mbox{max}}}_{a_{\tiny \mbox{min}}} da \frac{1}{\nh} \frac{d\ngr}{da} 4\pi\omega^2 f_a(\omega) 2\pi \frac{2}{3}\frac{\mu^2_\perp \omega^4}{c^3}
\end{equation} where \amin\ and \amax\ are the minimum and maximum size of the dust grain, and $\frac{1}{\nh} \frac{d\ngr}{da}$ is the number of dust grains per unit size per H atom.

The angular velocity distribution function for grain size $a$, $f_a(\omega)$, can be obtained by solving a stationary Fokker--Planck equation \citep{spdust2009},
\begin{equation}\label{eq:FP}
    \frac{df_a(\omega)}{d\omega}+\left[\frac{I\omega}{kT}\frac{F}{G} + \frac{\tau_{\tiny{\mbox{H}}}}{\tau_{\tiny{\mbox{ed}}}}\frac{1}{3G}\frac{I^2\omega^3}{(kT)^2}\right] f_a(\omega) = 0
\end{equation}
where $I$ is the moment of inertia of the dust particles, $\tau_{\tiny{\mbox{H}}}$ is the characteristic rotational damping time-scale for collisions with neutral H atoms, and $\tau_{\tiny{\mbox{ed}}}$ is the characteristic damping time scale for electric dipole radiation. Here $\tau_{\tiny{\mbox{H}}}$ and $\tau_{\tiny{\mbox{ed}}}$ are 
\begin{eqnarray}\label{eq:tscale}
    \tau_{\tiny{\mbox{H}}} & = & \left[\nh m_{\tiny \mbox{H}} \left(\frac{2kT}{\pi m_{\tiny \mbox{H}}}\right)^{1/2}\frac{4\pi a^4_{\tiny \mbox{cx}}}{3I} \right]^{-1} \\
    \tau_{\tiny{\mbox{ed}}} & = & \frac{I^2c^3}{2\mu_\perp^2kT}
\end{eqnarray}
where $a_{\tiny \mbox{cx}}$ is the ``cylinderical excitation-equivalent'' radius, defined as $4\pi a^4_{\tiny \mbox{cx}} \equiv \frac{3}{2}\oint\rho^2d\mbox{S}$ \citep{spdust2009}. 

In Equation~\ref{eq:FP}, $F$ and $G$ are the dimensionless damping and excitation coefficients of spinning dust particles for various interaction processes (e.g., collisions with ions and neutrals, plasma drag, infrared radiation, photoelectric emission, H$_2$ formation) and need to be computed for each interaction process \citep[e.g.,][]{draine_and_lazarian_1998,spdust2009}. 

Although the small grains may be sheetlike \citep{draine_and_lazarian_1998,spdust2009}, we adopt a spherical geometry for the grains when computing the damping and excitation coefficients for the impact of C-shocks, as many previous works do when computing $F$ and $G$ \citep[e.g.,][]{draine_and_lazarian_1998,spdust2009,silsbee_etal_2011,hoang_etal_2019}.

For computing \jvsp, we use the publicly available code \texttt{SpDust}\citep{spdust2009,silsbee_etal_2011} that solves Equation~\ref{eq:FP} numerically to obtain $f(\omega)$ and computes $\frac{\jvsp}{\nh}$ for an assumed grain size distribution $\frac{1}{\nh} \frac{d\ngr}{da}$, a dipole moment $\mu$ and ISM environment parameters (\nh, $T$, $x_{\mbox{\tiny H}}$, $x_{\mbox{\tiny M}}$ and $y$ in Table~\ref{tab:ismparam}).

\subsection{ISM Associated with Star-forming Regions}\label{sec:sfregion}
\begin{figure}[t]
\plotone{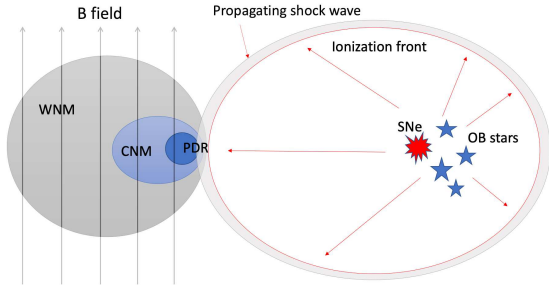}
\caption{Illustration of the star-forming region including \HII\ region and associated `magnetized' molecular cloud with the WNM, CNM, and dense PDR, where the potential AME can be observed before the `C-type' shock wave after the supernova explosion arrives and impacts the dust particles in the clouds.  
\label{fig:sfregion}}
\end{figure}

The classical multiphase model of the ISM proposed by \cite{mckee_and_ostriker_1977} is a simplified but still relevant picture of the ISM in star-forming regions. 
In Figure~\ref{fig:sfregion}, we show an illustrative picture of a star-forming region where the ``magnetized'' molecular cloud is associated with an \HII\ region. The molecular cloud consists of a multiphase ISM (WNM and CNM) and is exposed to UV radiation from the nearby \HII\ region where the ISM is fully ionized at $T\approx10^4$K by OB stars. A small part of the high-density molecular cloud forms a PDR where dust is present in abundance as both large grains and PAHs, as shown by the far-IR/submillimeter emission (from large grains) and near-IR emission (e.g. 8 $\mu$m from PAHs). A low-velocity shock wave propagating into the ``magnetized'' molecular cloud far from the \HII\ region creates a ``C-type'' shock. 

In this study, we assume that a high-velocity strong supernova shock wave (J-type shock) or strong UV radiation destroys the dust grains (including AME carriers) in the \HII\ region; therefore, AME does not rise from the \HII\ region itself, and the radio SED from the \HII\ region is dominated by free-free emission. We study the condition for AME formation in the CNM, WNM, and PDR in molecular clouds and investigate the relative strength of AME compared with free-free and thermal dust emissions that are also from the molecular cloud. 

The approximate volume filling factor $f_{\tiny V}$ for hot ionized medium (HIM), warm ionized medium (WIM), WNM and CNM in our Galaxy is 0.5, 0.1, 0.4, and 0.01, respectively \citep{draine_2011book}. Although the WNM fills a significant fraction of the volume of the galactic disk, the effective volume densities $n_{\tiny \mbox{H}} f_{\tiny V}$ of WNM and CNM are similar (0.2 and 0.3 \cmvol); therefore, it is likely that the contribution to the observed AME from CNM and WNM is similarly important for a given observing beam. 

If the OB stars are forming and creating the \HII\ region, the clouds are exposed to strong UV radiation, and some of them form a PDR. The significance of AME from a PDR with high density ($n_{\tiny \mbox{H}}\approx 10^5 \cmvol$) is difficult to assess due to its uncertain and probably very small volume filling factor; however, it could be an important contribution when the next-generation Very Large Array \citep[ngVLA;][]{ngvlabook} starts to resolve individual star-forming regions in galaxies with high angular resolution (0.5-50 mas).

Within a few tens of megayears, the supernova explodes and the shock wave propagates into the clouds. For a low-ionization medium with average or higher density like CNM, WNM, and PDR, the shock wave is a C-type where the shock velocities of the ionized and neutral particles are different due to the existence of a magnetic field, and therefore the neutral particle can have a supersonic drift velocity relative to the ionized particles \citep{draine_2011book}. 

\begin{figure}[t]
\plotone{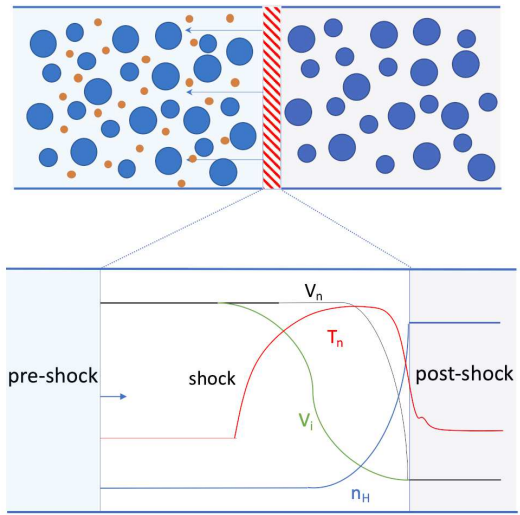}
\caption{Schematic view of the dust particles in gas clouds before and after the impact of the C-shock. The upper panel illustrates the dust particles of different sizes. The zoomed-in view of the shock layer (hatched by red lines) in the lower panel illustrates the velocity, temperature, and density profile of the gas experiencing magnetized shock creating different velocities for neutrals and ions ($V_n$ and $V_i$). The small orange circles seen in the pre-shock region represent the nanoparticles as an AME carrier that are broken up in the post-shock region by centrifugal force due to the spin-up by the collision with neutral particles drifting with supersonic drift velocity ($V_n-V_i$).
\label{fig:shocktube}}
\end{figure}

In Figure~\ref{fig:shocktube}, we show a schematic picture of the propagation of a C-shock into the ISM with dust particles having different sizes. The small orange circles in the pre-shock region on the left-hand side of the shock layer (hatched red lines in the middle) represent the nanoparticles as a source of AME, and the other large dust grains existing in both pre- and post-shock regions are represented by larger blue circles. The velocity, temperature, and density profile of the ISM in the shock layer are shown in the zoom-in panel. If a charged dust grain sticks with the magnetic field when the C-shock propagates (to the left), the neutral particles with high drift velocity ($V_n-V_i$) relative to the ions collide with the dust particles via stochastic bombardment and introduce additional damping and excitation terms, which destroys the small dust grains (small orange circles in the pre-shock region). We will discuss the process in detail in Sections~\ref{sec:dynamics} and~\ref{sec:destroy}.

\begin{deluxetable}{lccc}[ht!]
\centering
\tablecaption{ISM Environment Parameters from \cite{draine_and_lazarian_1998}}\label{tab:ismparam}
\tablehead{\colhead{Parameter} &
           \colhead{CNM}       &
           \colhead{WNM}       &
           \colhead{PDR}       }
\startdata
\nh [\cmvol]  &  30 & 0.4 & $10^5$\\
$T$ [K] & 100 & 6000 & 300\\
$T_d$ [K] & 20  & 20 & 50\\
$\chi$ & 1 & 1 & 3000\\
$x_{\mbox{\tiny H}}\equiv n(\mbox{H}^+)/$\nh & 0.0012 & 0.1 & 0.0001\\
$x_{\mbox{\tiny M}}\equiv n(\mbox{M}^+)/$\nh & 0.0003 & 0.0003 & 0.0002\\
$y\equiv 2n(\mbox{H}_2)/$\nh & 0 & 0 & 0.5\\
\enddata
\end{deluxetable}

\section{Rotational Dynamics of Spinning Nanoparticles in C-shocks} \label{sec:dynamics}
In the ISM associated with star-forming regions (CNM, WNM, and PDR) from which we expect to observe both free-free emission and AME, a large abundance of small nanoparticles with a high spin angular momentum resulting from the shattering of large grains by supernova winds \citep{jones_etal_1996} is expected to increase the strength of AME. However, on the other hand, one can also expect that the high spin angular velocity of the nanoparticles creates a strong centrifugal force disrupting the particles themselves. This rotational disruption can decrease the abundance of the smallest nanoparticles, which can decrease the resulting AME. 

Recently, \cite{hoang_etal_2019} introduced a model of spinning dust in C-shocks, accounting for this destruction effect. We will incorporate this process in the \texttt{SpDust} code \citep{spdust2009,silsbee_etal_2011} and compute the emissivity of spinning nanoparticles in the presence of C-shocks.     

\subsection{Structure of C-shocks in CNM, WNM, and PDR}
Shocks in molecular clouds where H$_2$ rovibrational cooling is efficient are common and expected to often be C-type \citep{draine_2011book}. The energy dissipation in C-shocks with two fluids (ions and neutrals) is continuous rather than impulsive, and efficient radiative cooling keeps the gas cool \citep{draine_2011book}.   
We use the Paris--Durham shock code \citep{flower_etal_2003,lesaffre_etal_2013,godard_etal_2019} to compute the velocity profiles of neutral and ion particles in C-shocks for the ISM with typical environmental parameters (Table~\ref{tab:ismparam}) for CNM, WNM, and PDR \citep{draine_and_lazarian_1998}.

The simulated shock is static. We assume that ISM parameters like temperature and density are continuous and not drastically different in the pre- and post-shock region because of the nature of the continuity of C-shocks. For the pre-shock parameters, we use the default values in the code,  except for hydrogen density, gas temperature, and dust temperature ($n_{\tiny \mbox{H}}$, $T$, and $T_d$) which are chosen for the CNM, WNM, and PDR from Table~\ref{tab:ismparam}. For the initial shock velocity, we note that the velocity of C-type shocks is low (5-25km/s) in general, and the high-velocity shocks are J-type \citep{godard_etal_2019}. Therefore, we choose 20km/s for the initial shock velocity. For the magnetic field strength, the Paris--Durham shock code uses the following parameterization \citep{godard_etal_2019}:

\begin{equation}\label{eq:bcode}
    B=b\sqrt{\frac{\nh}{1cm^{-3}}} ~\mu \mbox{G}
\end{equation} where $b=1$ by default in the shock code.

\begin{figure*}
\gridline{\fig{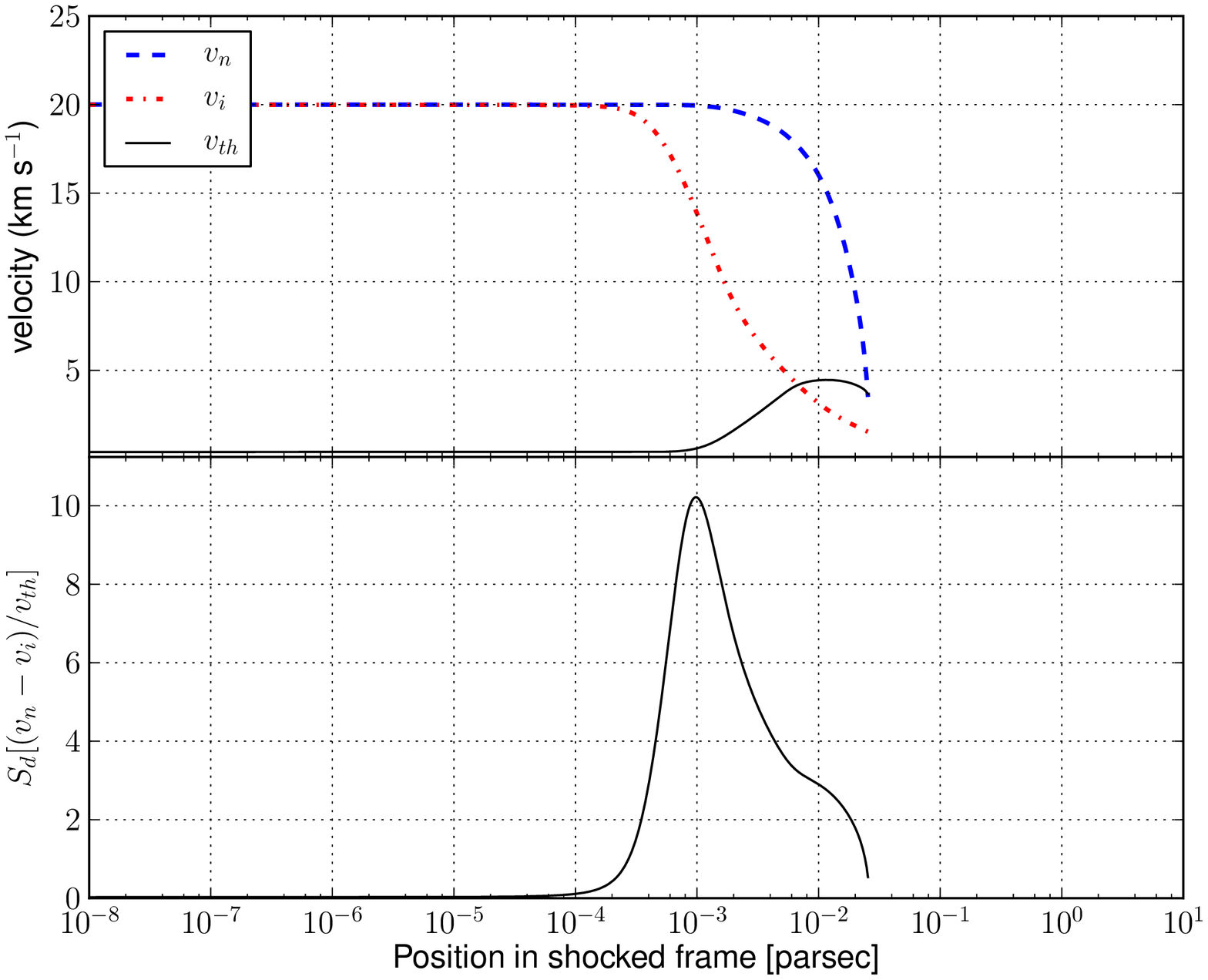}{0.5\textwidth}{(a) CNM}
          \fig{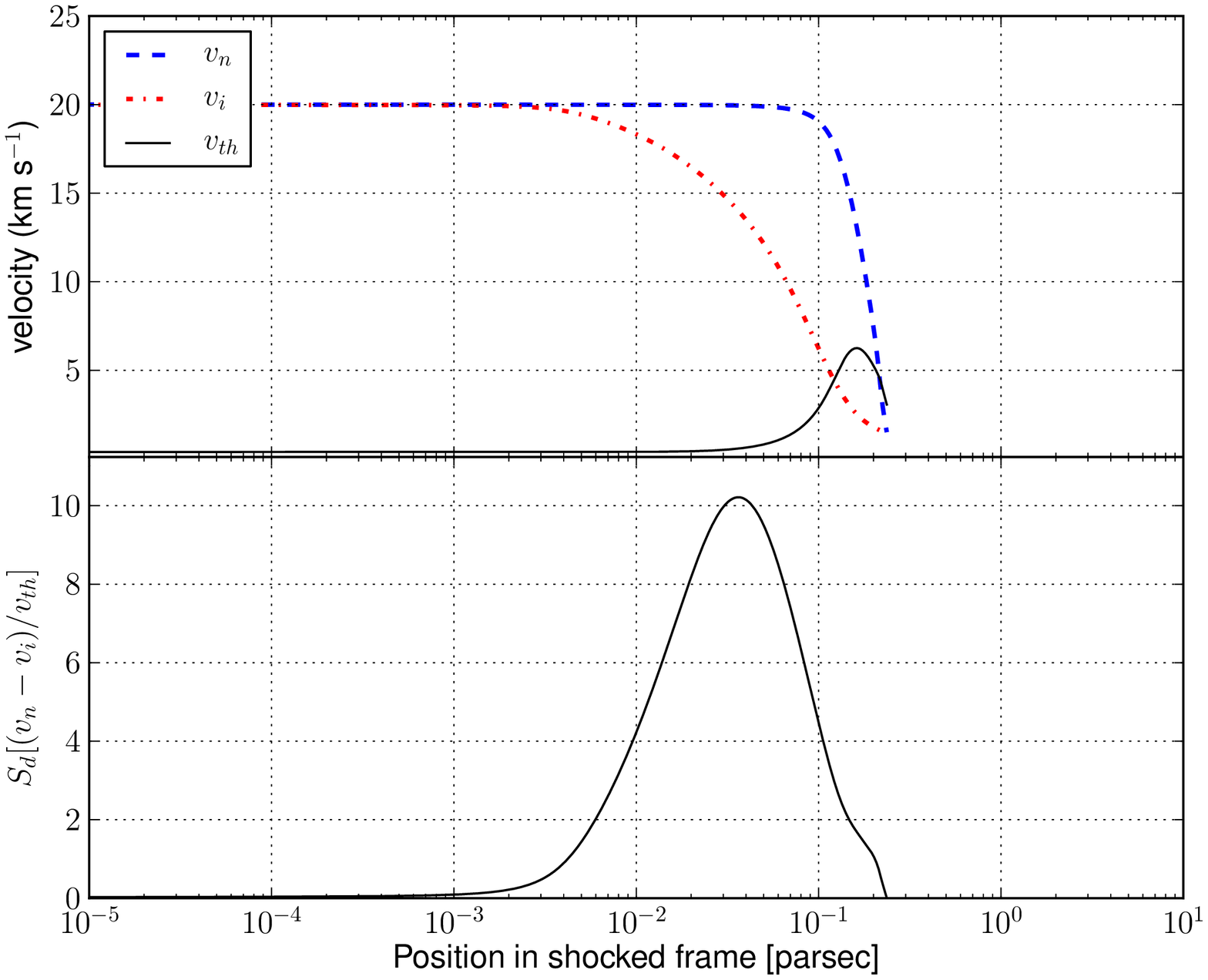}{0.5\textwidth}{(b) WNM}}
\gridline{\fig{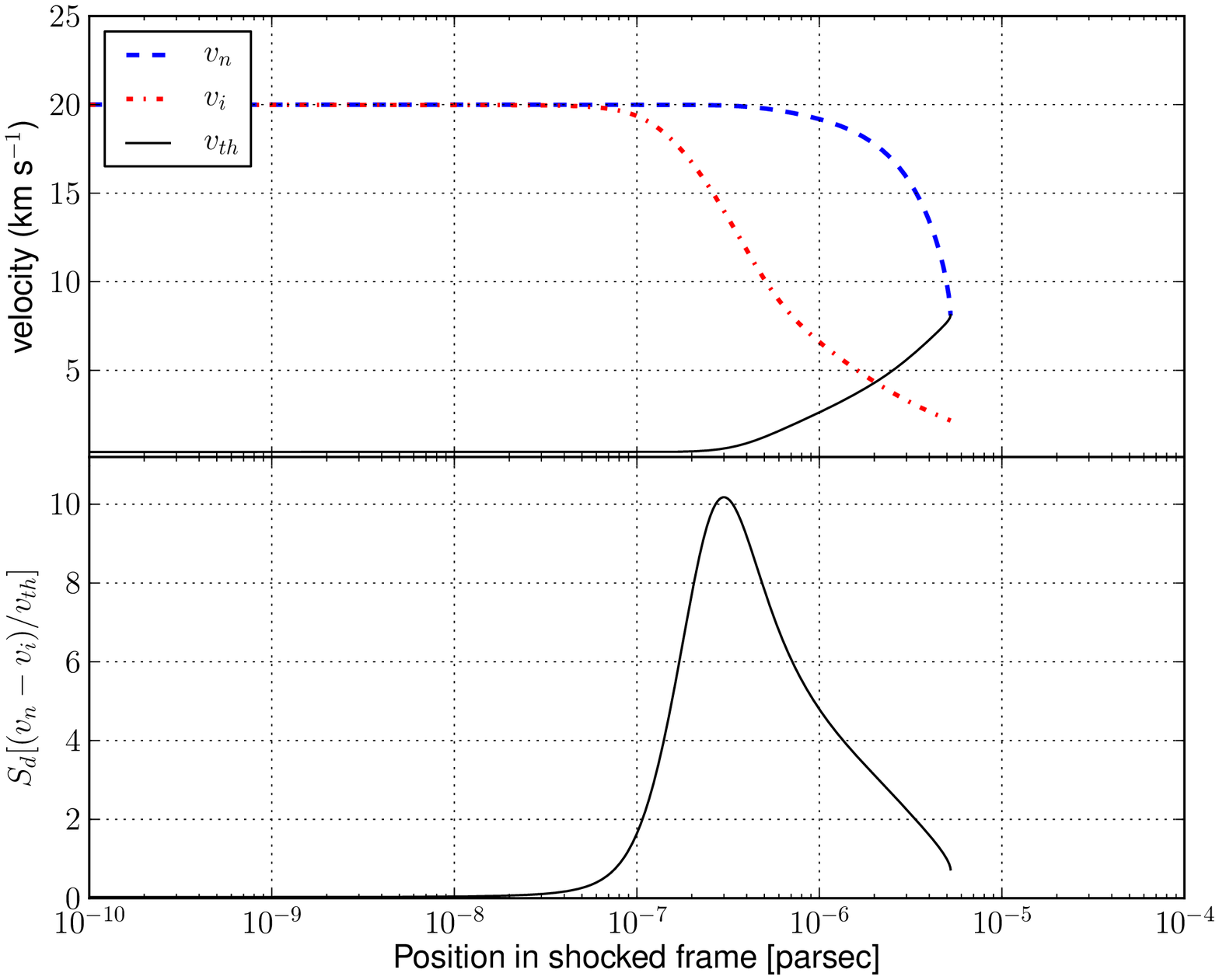}{0.5\textwidth}{(c) PDR}
}
\caption{\textit{Upper panels:} shocked-frame velocity profiles of neutrals (blue dashed line) and ions (red dotted-dashed line) with a thermal velocity profile (solid black line) \textit{Lower panels:} drift velocity of neutrals with respect to the thermal velocity, $s_d=\frac{v_n-v_i}{v_{th}}$. The initial shock velocity (20 km/s) is the same for all cases.}\label{fig:shockprofile}
\end{figure*}

In the upper panels of the plots in Figure~\ref{fig:shockprofile}, we show the velocity profiles of neutrals ($v_n$, blue dashed line) and ions ($v_i$, red dotted-dashed line) in the shock frame for C-shocks in the magnetized ISM, for the physical conditions of CNM, WNM, and PDR in Table~\ref{tab:ismparam} and a 20 km/s shock velocity. The solid black line is the thermal velocity profile ($v_{th}$). In the lower panels of the plots in Figure~\ref{fig:shockprofile}, we show the profile of the ratio between drift velocity ($v_n-v_i$) and thermal velocity ($v_{th}$):

\begin{equation}
    s_d=\frac{v_n-v_i}{v_{th}}
\end{equation}

Although the detailed shapes of the profile vary depending on the pre-shock density, temperature, and strength of irradiated radiation \citep[e.g.,][]{godard_etal_2019}, the neutrals and ions have different velocity profiles for all three ISM conditions (CNM, WNM, and PDR) because ions are decelerated by the magnetic field, and the drift velocity becomes supersonic ($s_d\gg1$) at the shock layer (Figure~\ref{fig:shockprofile}). Note that the computation of the velocity profiles stops when the neutrals and ions are about to recouple; however a full profile is not necessary to confirm that the supersonic drift is developed in the C-shock. 

The Mach number of the neutral drift velocity, $s_d$ depends on the shock velocity and sound speed, and the C-shock develops supersonic drift for the typical values of shock velocity (a few tens of kilometers per second) found in the ISM velocity--density plane \citep[e.g.,][]{draine_mckee_1993}. The C-shock with supersonic drift ($s_d\gg1$) has an impact on the calculation of the damping and excitation coefficients ($F$ and $G$) described in the following section (Section~\ref{sec:rotdyn}).

\subsection{Rotational Damping and Excitation in C-shocks}\label{sec:rotdyn}
Various damping and excitation processes in the rotational dynamics of grain particles are captured by the dimensionless damping and excitation coefficients $F$ and $G$ in Equation~\ref{eq:FP}. In this work, we consider additional damping and excitation processes due to the presence of supersonic neutral drift relative to charged grains \citep{hoang_etal_2019}. 

As discussed in Section~\ref{sec:ame_instro}, we assume a spherical geometry when computing damping and excitation coefficients. For the damping and excitation coefficients for the supersonic ($s_d \gg 1$) and transonic cases ($s_d \sim 1$), we follow  \cite{roberge_etal_1995} and adopt their results, as \cite{hoang_etal_2019} does.

Let $\hat{\boldsymbol{x}}\hat{\boldsymbol{y}}\hat{\boldsymbol{z}}$ be the reference frame fixed to the gas, such that the $\hat{\boldsymbol{z}}$-axis is directed along the magnetic field, and the drift velocity $v_d$ lies in the $\hat{\boldsymbol{y}}\hat{\boldsymbol{z}}$-plane with an angle $\alpha$ with $\hat{\boldsymbol{z}}$. We consider a perpendicular shock ($\alpha=90^\circ$). 

From \cite{roberge_etal_1995}, the dimensionless damping coefficients $\langle\Delta j_i\rangle$ are
\begin{equation}\label{eq:F}
\langle\Delta j_i\rangle = 
   \begin{cases}
     -\left[\delta + M_0(s_d)\right]j_i, & i=x,y, \\
     -M_0(s_d)j_i, & i=z.
  \end{cases}
\end{equation} 
In this equation, 
\begin{equation}
    M_0(s_d) = \frac{\sqrt{\pi}}{4s_d}\left[2(1+s^2_d)\mbox{erf}(s_d) - P\left(\frac{3}{2},s^2_d\right)\right]
\end{equation} where $P$ is an incomplete gamma function, and $\delta$ is a magnetic damping parameter measuring the relative efficiency of magnetic versus gas damping, in the sense that a large $\delta$ value would correspond to efficient magnetic alignment \citep{roberge_etal_1995}, and is written as \citep{roberge_etal_1993}
\begin{equation}
    \delta = \frac{3KVB^2}{4\sqrt{\pi}\nh m_{\tiny \mbox{H}}v_{th}b^4\Gamma_{\|}}.
\end{equation} where $K=10^{-13}\left(\frac{T_d}{18\mbox{\tiny K}}\right)^{-1}$ \citep{draine_2011book}, $V=\frac{4\pi}{3}a_{\mbox{\tiny cx}}^3$, $v_{th} = \sqrt{\frac{2kT}{m_{\tiny \mbox{H}}}}$, $b\approx a_{\mbox{\tiny cx}}$, and $\Gamma_{\|}=1.0$ for spherical grains \citep{roberge_etal_1993}.
For the magnetic field strength ($B$), we use Equation~\ref{eq:bcode}. Although $\delta$ is not well constrained due to the highly uncertain material parameter $K$\citep{roberge_etal_1993}, we find that $\delta \ll M_0(s_d)$ for the ISM condition that we are considering (i.e., high temperature in the C-shock region). 

The dimensionless angular momentum $j_i (i=x,y,z)$ is $j_i\equiv\frac{J_i}{\sqrt{IkT}}$ and $j_i\approx1$ for a spherical grain rotating with the equipartition energy \citep{roberge_etal_1995}. The dimensionless damping coefficient $F_{sd}$ for spherical grains due to supersonic neutral drift in the C-shock becomes
\begin{equation} 
F_{sd} = \frac{1}{3}\displaystyle\sum_{i=x,y,z} \langle\Delta j_i\rangle
\end{equation}

Also from \cite{roberge_etal_1995}, the dimensionless excitation coefficients $\langle (\Delta j_i)^2 \rangle$ are 
\begin{equation}
    \langle (\Delta j_x)^2 \rangle = D_T + \frac{T_d}{T} \left[2\delta+M_0(s_d)\right]\\
\end{equation}
\begin{equation}
    \langle (\Delta j_y)^2 \rangle = D_P \sin^2\alpha + D_T \cos^2\alpha +  \frac{T_d}{T} \left[2\delta+M_0(s_d)\right]\\
\end{equation}
\begin{equation}
    \langle (\Delta j_z)^2 \rangle = D_P \cos^2\alpha + D_T \sin^2\alpha +  \frac{T_d}{T} M_0(s_d)\\
\end{equation}
The quantities 
\begin{equation}
    D_T(s_d)=\frac{3}{4}\left[\left(1+2s_d^2\right) M_0(s_d) + \left( 1-2s_d^2 \right) M_2(s_d)\right]
\end{equation}
and 
\begin{equation}
    D_P(s_d)=\frac{3}{2}\left[M_0(s_d)-M_2(s_d)\right]
\end{equation} where
\begin{equation}
\begin{aligned}
    M_2(s_d) = \frac{\sqrt{\pi}}{4}s_d\mbox{erf}(s_d) - \frac{3\sqrt{\pi}}{16}s_d^{-3}P\left(5/2,s_d^2\right) \\ + \frac{\sqrt{\pi}}{4}s_d^{-3}P\left(3/2,s_d^2\right)
\end{aligned}
\end{equation}
are dimensionless, monotonically increasing functions of $s_d$ that have the limiting values $D_T(0)=D_P(0)=1$ and satisfy the inequality $D_T(s_d) > D_P(s_d)$ for $s_d>0$ \citep{roberge_etal_1995}. The dimensionless excitation coefficient $G_{sd}$ for spherical grains due to supersonic neutral drift in C-shocks becomes
\begin{equation} 
G_{sd} = \frac{1}{3}\displaystyle\sum_{i=x,y,z} \langle \left(\Delta j_i\right)^2\rangle
\end{equation}.

\section{Rotational Disruption of Spinning Nanoparticles}\label{sec:destroy}
\subsection{Critical Angular Velocity for Disruption}
The tensile stress of a spherical grain with mass $m$, radius $r$ and angular velocity $\omega$ is $S = \frac{1}{4}\rho\omega^2 a^2$, which is converted to the relation between the critical angular velocity for grain disruption and the maximum grain tensile strength, $S_{\mbox{\tiny max}}$ \citep{hoang_2020}, 
\begin{equation}\label{eq:wcri}
    \frac{\omega_{cri}}{2\pi} \simeq 5.72\times10^{10} a^{-1}_{-7} S^{1/2}_{\mbox{\tiny max,9}} \hat{\rho}^{-1/2}~~[\mbox{Hz}]
\end{equation} where $a_{-7}=a/(10^{-7}\mbox{cm})$, $S_{\mbox{\tiny max,9}}=S_{\mbox{\tiny max}}/(10^{9}~\mbox{erg cm}^{-3})$ and $\hat{\rho}=\rho/(3~\mbox{g cm}^{-3})$.

The exact value of the maximum tensile strength $S_{\mbox{\tiny max}}$ depends on the composition and structure of the dust grain and is largely unknown; compact grains can have higher tensile strength than porous/composite grains \citep[e.g.,][]{hoang_etal_2019}. For example, a polycrystalline bulk solid has $S_{\mbox{\tiny max}}\sim10^9-10^{10}$erg cm$^{-3}$ while ideal material, like diamond, can have $S_{\mbox{\tiny max}}\gtrsim10^{11}$erg cm$^{-3}$\citep[][]{hoang_2020}. Since the smallest grains may be sheetlike and have approximately 100 carbon atoms corresponding to the size of large PAH \citep{draine_and_lazarian_1998,spdust2009}, it is a reasonable assumption that the tensile strength of the nanoparticles considered in this study is not very large because they are not compact. Therefore, in this study, we consider $S_{\mbox{\tiny max}}$ to be varied for a range, $10^8$--$10^{10}$erg cm$^{-3}$. However, we emphasize that the result of the grain destruction model in this study depends on the poorly understood maximum tensile strength parameter.

\subsection{Spin Angular Velocity of Nanoparticles}
If one assumes Maxwellian distribution for $f_a(\omega)$ with gas temperature $T$, one can solve Equation~\ref{eq:FP} to obtain $\langle\omega^2\rangle$ \citep{draine_and_lazarian_1998}:
\begin{equation}\label{eq:wrot}
   \langle\omega^2\rangle = \frac{2}{1+\left[1+(G/F^2)(20\tau_{\tiny{\mbox{H}}}/\tau_{\tiny{\mbox{ed}}})\right]^{1/2}}\left(\frac{G}{F}\right)\left(\frac{3kT}{I}\right) 
\end{equation}
This analytic expression lets us compute the angular velocity $\omega_{rot} \equiv \sqrt{\langle\omega^2\rangle}$ of grain particles with a rotational temperature $T_{rot}$ for a given grain size $a$ \citep{hoang_etal_2019},
\begin{eqnarray}\label{eq:wrot2}
    \frac{\omega_{rot}}{2\pi} & = & \frac{1}{2\pi}\left(\frac{3kT_{rot}}{I}\right)^{1/2} \nonumber \\
    & \simeq & 1.4\times 10^{10} a^{-5/2}_{-7} \left(\frac{T_{rot}}{10^3\mbox{K}}\right)^{1/2} \hat{\rho}^{-1/2}~[\mbox{Hz}]
\end{eqnarray}
which will be compared to the critical angular velocity $\omega_{cri}$ in Equation~\ref{eq:wcri} for centrifugal disruption of grain particles. The rotational temperature $T_{rot}$ is related to the gas temperature $T$ by
\begin{equation}
    \frac{T_{rot}}{T} = \frac{2}{1+\left[1+(G/F^2)(20\tau_{\tiny{\mbox{H}}}/\tau_{\tiny{\mbox{ed}}})\right]^{1/2}}\left(\frac{G}{F}\right)
\end{equation} and if the nanoparticles are in suprathermal rotation, $T_{rot}>T$.

\begin{figure}[ht!]
\plotone{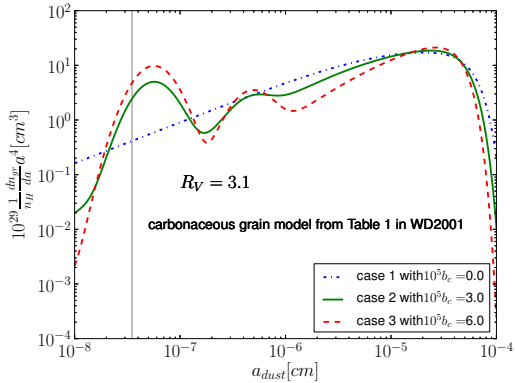}
\caption{Grain volume per H atom per logarithmic interval in dust size, $(4\pi a^3/3)d\ngr/d\mbox{\small ln}a$. Blue dotted-dashed, green solid, and red dashed lines are the models with parameters in lines 1, 4, and 7 of Table 1 in \cite{weingartner_and_draine_2001}, labeled as ``case1'', ``case2'', and ``case3'', respectively, for subsequent analysis. The gray vertical line indicates the minimum dust size, $a_{\mbox{\tiny min}}=3.5$\AA, used in \texttt{SpDust}.}
\label{fig:adist}
\end{figure}

\subsection{Grain size distribution}
The emissivity of AME is determined by the ensemble of electric dipole emission from nanoparticles with a distribution of their size $a$ ($\frac{1}{\nh} \frac{d\ngr}{da}$ in Equation~\ref{eq:emissi}), and most of the AME emissivity is from the smallest particle ($a_{-7}\lesssim0.5$), as shown in \cite{spdust2009}. Therefore, the shape of the grain size distribution, especially for small grains, is a very important component of the AME models \citep{hensley_and_draine_2017}; however, is not well understood for different ISM environments. Therefore, for carbonaceous grains, we assume that the grain size distribution follows the commonly used form of the composite of log-normal and power-law distribution by \cite{weingartner_and_draine_2001} for grain radii $a_{\mbox{\tiny min}}=3.5$\AA\ $<a<a_{\mbox{\tiny max}}=100$\AA\ as adopted in the \texttt{SpDust} code \citep{spdust2009},
\begin{equation}\label{eq:asize}
\begin{aligned}
   \frac{1}{\nh} \frac{d\ngr}{da} = D(a) + \frac{C}{a}\left(\frac{a}{a_t}\right)^{\alpha} F(a;\beta;a_t)\\
   \times 
   \begin{cases}
     1 & a_{\mbox{\tiny min}}<a<a_t, \\
     e^{-\left[(a-a_t)/a_c\right]^3}& a>a_t
  \end{cases}
\end{aligned}
\end{equation}
where 
\begin{equation}
    F(a;\beta;a_t)=
    \begin{cases}
     1+\beta\frac{a}{a_t} & \beta \geq 0, \\
     \left(1-\beta\frac{a}{a_t}\right)^{-1} & \beta < 0
  \end{cases}
\end{equation}
and the log-normal distribution $D(a)$ is
\begin{equation}  
    D(a)=\displaystyle\sum^{2}_{i=1}\frac{B_i}{a}\mbox{exp}\left\{ -\frac{1}{2}\left[\frac{\mbox{ln}(a/a_{0,i})}{\sigma}\right]^2\right\}.
\end{equation} with the normalization $B_i$ defined to place a total number $b_{C,i}$ of carbon atoms per H atom in the $i$th log-normal distribution. Here $b_{C,1}=0.75b_C$ and $b_{C,2}=0.25b_C$, $b_C$ being the total carbon abundance per H atom, $a_{0,1}=3.5$\angst, $a_{0,2}=30$\angst, and $\sigma=0.4$. This size distribution has a total of six parameters ($b_C,C,a_t,a_c,\alpha$, and $\beta$). 

In Figure~\ref{fig:adist}, we plot the grain volume per H atom per logarithmic interval in dust size, $(4\pi a^3/3)d\ngr/d\mbox{\small ln}a$ for three different models in \cite{weingartner_and_draine_2001} that matches the extinction curve of the diffuse ISM. Blue dotted-dashed, green solid, and red dashed lines are the models with parameters in lines 1, 4, and 7 of Table 1 in \cite{weingartner_and_draine_2001}, and the gray vertical line indicates the minimum dust size, $a_{\mbox{\tiny min}}=3.5$\AA, used in \texttt{SpDust}. In Figure~\ref{fig:adist}, we label each model as ``case1'', ``case2'', and ``case3'', respectively.

The exact shape of the size distribution of the dust grains varies depending on the ISM condition and is not well known for the ISM in star-forming regions exposed to extreme physical conditions (strong radiation and shock). In this study, we use the three labeled models in Figure~\ref{fig:adist} to investigate a range of the dust size distribution from the case without additional enhancement of small grains to the case with significant enhancement of small grains as represented by the log-normal component, which leads to stronger AME. In principle, strong UV radiation from a massive star-forming region can exert radiation torque on dust grains, and the fraction of small grains can increase by disruption of large grains \citep{hoang_etal_rat_2019}, which then may increase the AME in an early stage of massive star formation before the C-shock impacts the ISM and destroys the small nanoparticles. However, the grain size distribution is poorly constrained, and extreme ISRF may be required to destroy large grains (see the discussion in Section~\ref{sec:riseame}).

\section{Radio-Millimeter Emission from Star-forming Regions}\label{sec:emission}
Recent observations of AME from star-forming regions challenge the widely accepted notion about the radio-millimeter SED; the radio-millimeter SED from star-forming regions in the frequency range $\approx 10$--$100$ GHz is dominated by free-free emission and thermal dust emission. Since the observed SED is based on flux measurements with a finite beam, unknown ``environmental'' factors, such as the geometry and volume filling factor of the ISM that are not directly related to the physics of the emission, affect the SED. Therefore, we only focus on the emissivity of each radiation process in order to isolate the environmental factors and characterize the significance of AME relative to the free-free and thermal dust emission.

Originally, the emissivity $j_{\nu}$ is defined as energy emitted at a frequency $\nu$ per unit volume, time, and solid angle (erg s$^{-1}$sr$^{-1}$cm$^{-3}$ Hz$^{-1}$). However, in this work, we use the emissivity per H atom $\frac{j_\nu}{\nh}$ in erg s$^{-1}$ sr$^{-1}$ Hz$^{-1}$ (H atom)$^{-1}$ (see Equation~\ref{eq:emissi}) following previous works \citep{draine_and_lazarian_1998,spdust2009}. The emissivities of free-free and thermal dust emission are computed as a function of frequency $\nu$ for CNM, WNM, and PDR ISM conditions (Table~\ref{tab:ismparam}) and the emissivity of AME (Equation~\ref{eq:emissi}) before and after the impact of the C-shock will be computed by following the process in Section~\ref{sec:dynamics} and compared with the emissivity of free-free and thermal dust emission.

\subsection{Calculating the Emissivity of Free-Free, Thermal Dust Emission and AME}
\subsubsection{Free-Free emission}
The emissivity per H atom of free-free emission with electrons of temperature $T$ is \citep{radibook} 
\begin{equation}\label{eq:jvff}
    \frac{\jvff}{\nh} = \frac{1}{4\pi}2^5\pi\left(\frac{e^6}{3m_e c^3}\right)\left(\frac{2\pi}{3k m_e}\right)^{1/2}g_{ff}\frac{\nh}{\sqrt{T}}e^{-\frac{h\nu}{kT}}
\end{equation} where $g_{ff}$ is the Gaunt factor for the free-free transition. The free-free emissivity is implemented in \texttt{SpDust}\citep{spdust2009,silsbee_etal_2011} using $g_{ff}$ tabulated from \cite{sutherland_1998}.

\subsubsection{Thermal dust emission}
The emissivity per H atom of thermal dust emission with dust temperature $T_d$ is \citep{radibook}
\begin{equation}\label{eq:jvbb1}
\begin{aligned}
    \frac{\jvbb}{\nh} = \frac{\kappa_0}{\nh}\left(\frac{\nu}{\nu_0}\right)^{\beta}\frac{2h\nu^3}{c^2}\frac{1}{e^{\frac{h\nu}{k T_d}}-1}
\end{aligned}
\end{equation}
where $\kappa_0$ is the dust volume absorption coefficient (i.e., cross section per unit volume), and $\beta$ is the emissivity spectral index. Instead of $\kappa_0$, often the more frequently used coefficient is the mass absorption coefficient  $\kappa_0^{\prime}$ (cm$^2$/g) and $\kappa_0=\kappa_0^{\prime}\rho_d$ for dust mass density $\rho_d$ (total dust mass/ISM volume). If gas and dust are well mixed in the same volume, $\rho_d$ can be inferred by the gas mass density ($\rho_{\mbox{\tiny gas}}=\nh m_{\mbox{\tiny H}}$) and dust-to-gas mass ratio $M_d/M_g=0.0083\rho_{\mbox{\tiny ref}}/(3~\mbox{g cm}^{-3})$ \citep{draine_2011book},
\begin{equation}
\begin{aligned}
    \rho_d =\nh m_{\mbox{\tiny H}}\times0.0083\left(\frac{\rho_{\mbox{\tiny ref}}}{3~\mbox{g/cm}^{-3}}\right)
\end{aligned}
\end{equation} 
where the reference density $\rho_{\mbox{\tiny ref}}$ is the solid density of a grain particle and different for different types of dust particles with an intermediate value of $3~\mbox{g/cm}^{-3}$ \citep{draine_2011book}. Both $\kappa_0^{\prime}$ and $\beta$ vary depending on the type of dust and wavelength. In this work, we adopt $\kappa_0^{\prime}=1.8$ (cm$^2$/g) at $\nu_0=599.98$ GHz from \cite{clark_etal_2016} and $\beta=2.0$ \citep[e.g.,][]{schnee_etal_2010}. Then \jvbb\ can be written as 
\begin{equation}\label{eq:jvbb}
\begin{aligned}
\frac{\jvbb}{\nh} = 0.015\left(\frac{\rho_d}{3 \mbox{g/cm}^{3}}\right)\frac{\nh m_{\mbox{\tiny H}}}{\nh} \left(\frac{\nu}{\nu_0}\right)^{2} \frac{2h\nu^3}{c^2}\frac{1}{e^{\frac{h\nu}{k T_d}}-1}
\end{aligned}
\end{equation}

We assume that even though the small nanoparticles (i.e., the source of AME) are destroyed by shocks, the bulk of the grain volume is still dominated by large grain particles (Figure~\ref{fig:adist}), and the thermal dust emission is not much affected by the disruption of the small nanoparticles. Therefore, we use the same \jvbb\ before and after the shock. We implemented the emissivity of thermal dust emission in the \texttt{SpDust} code.

\subsubsection{AME}
The emissivity per H atom of AME is shown in Equation~\ref{eq:emissi}. We modify the \texttt{SpDust} code and add the damping and excitation coefficients $F_{sd}$ and $G_{sd}$ due to the interaction with supersonic neutral drift in C-shocks (Section~\ref{sec:dynamics}) to the other damping and excitation coefficients, $F_i$ and $G_i$ when computing the total damping and excitation coefficients:
\begin{equation}
\begin{aligned}\label{eq:coeffs}
    F=\displaystyle\sum_{i}F_i + F_{sd}\\
    G=\displaystyle\sum_{i}G_i + G_{sd}.
\end{aligned}
\end{equation}

Since grains might not spin around the axis of their greatest inertia, \cite{silsbee_etal_2011} introduced the correction terms to the damping coefficient by incoming particles. We also introduce the correction terms to the damping coefficient, $F_{sd}$ in order to account for the randomized orientation of grains relative to their angular momentum vector. The impact of C-shocks is the collision between neutrals (H, H$_2$ and He) and charged grains, and we use the correction terms (Equation (140) in \cite{silsbee_etal_2011}) for the charged grain with neutral impactors.

\subsubsection{Total Emissivity in Radio-millimeter Wavelength}
Using \texttt{SpDust}, we compute the total emissivity per H atom for the emission in the radio-millimeter wavelength, including free-free, thermal dust, and AME emissivity, using Equation~\ref{eq:jvff}, ~\ref{eq:jvbb}, and ~\ref{eq:emissi}: 
\begin{equation}
\frac{j_\nu}{\nh} = 
    \left(\frac{\jvff}{\nh} + \frac{\jvbb}{\nh} + \frac{\jvsp}{\nh} \right)
\end{equation}
The original input parameters to run \texttt{SpDust} are the ISM environment parameters (Table~\ref{tab:ismparam}). Two additional parameters that we add to these standard \texttt{SpDust} parameters are the dust emissivity spectral index ($\beta$ in Equation~\ref{eq:jvbb1}) and the minimum size of nanoparticles ($a_{\mbox{\tiny min}}$ in Equation~\ref{eq:asize}).

The total emissivity of the radio-millimeter emission is computed for three different ISM environments, CNM, WNM, and PDR, before and after the C-shock impacts the ISM. 

\subsection{AME before the Impact of the C-shock}
\begin{figure*}
\gridline{\fig{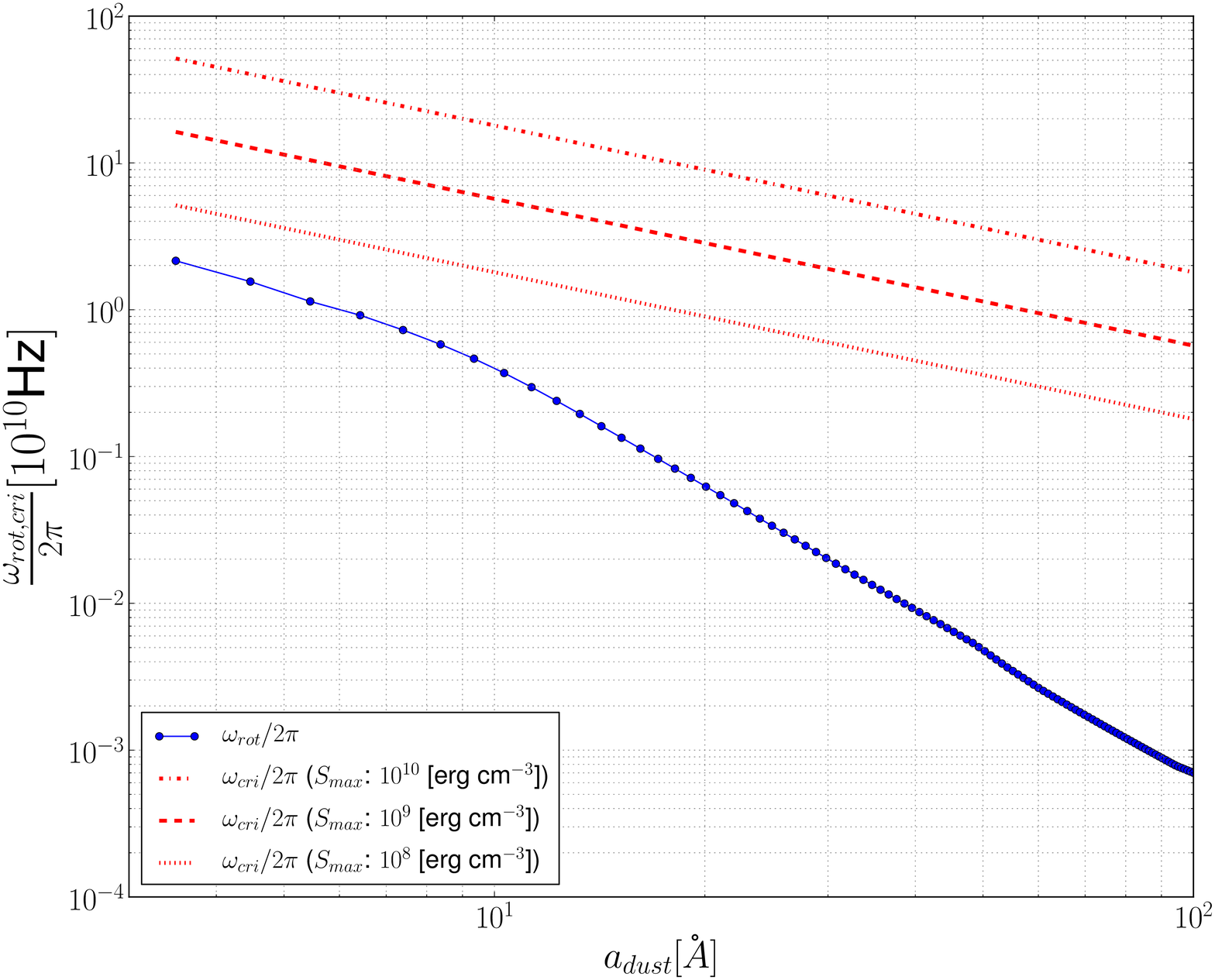}{0.5\textwidth}{(a) CNM}
          \fig{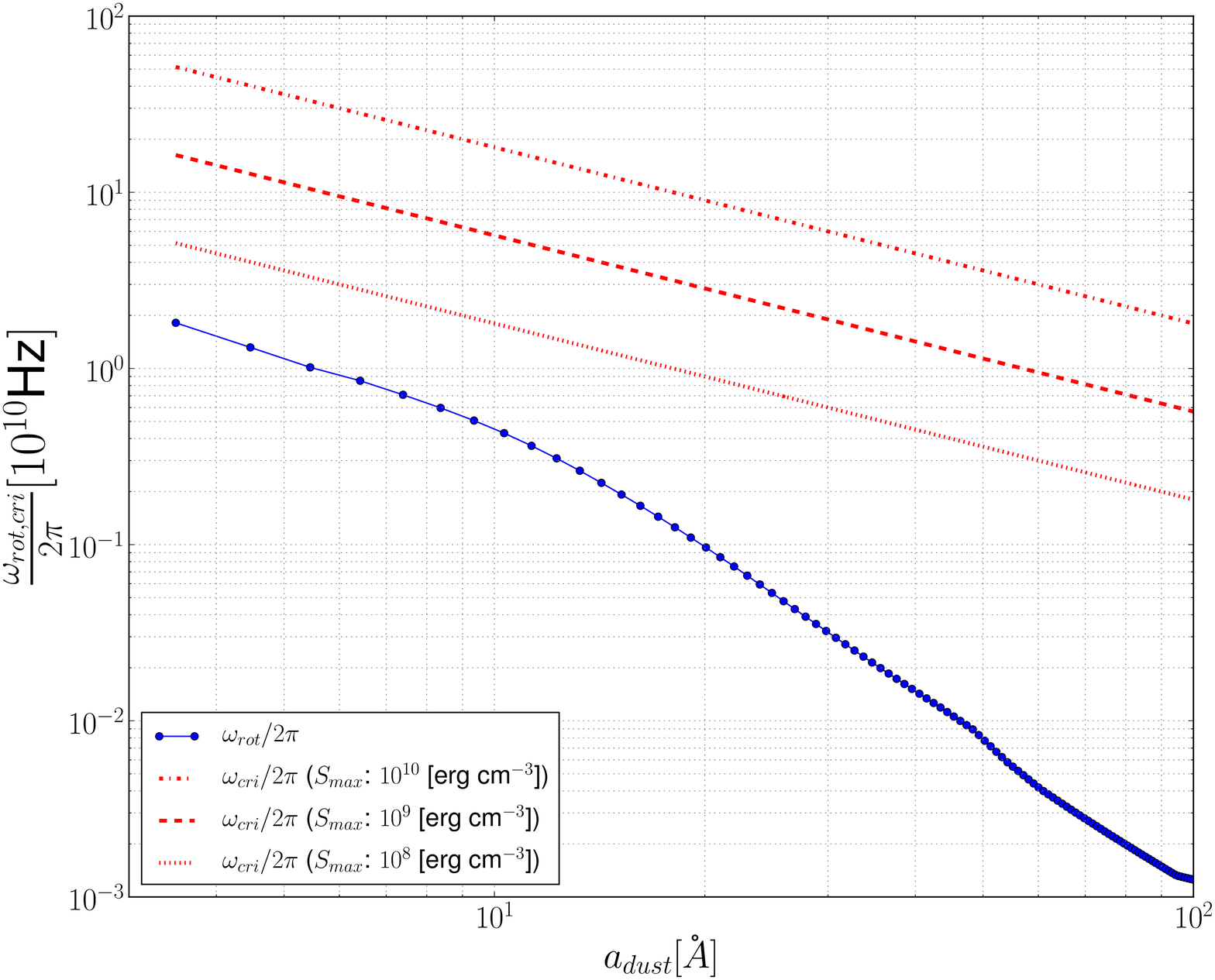}{0.5\textwidth}{(b) WNM}}
\gridline{          
          \fig{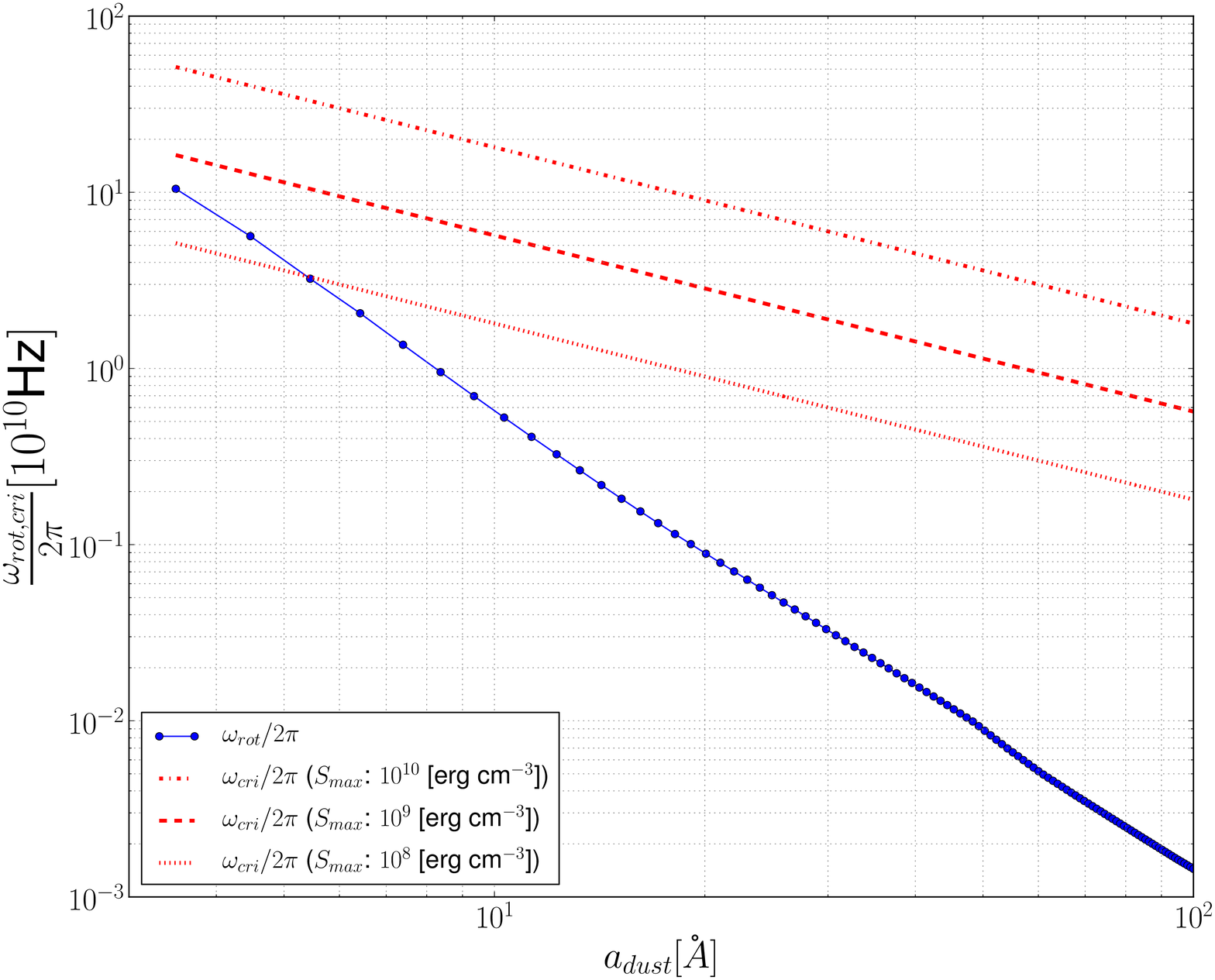}{0.5\textwidth}{(c) PDR}
}
\caption{\textit{Before the impact of the C-shock}: Angular rotation velocity of the dust particle (blue line) emitting AME and the critical angular rotation velocity (red lines) for a range of the maximum tensile strength ($10^8 - 10^{10}$ erg cm$^{-3}$) of the spinning dust particles} 
\label{fig:wrot_before}
\end{figure*}
\begin{figure*}
\gridline{\fig{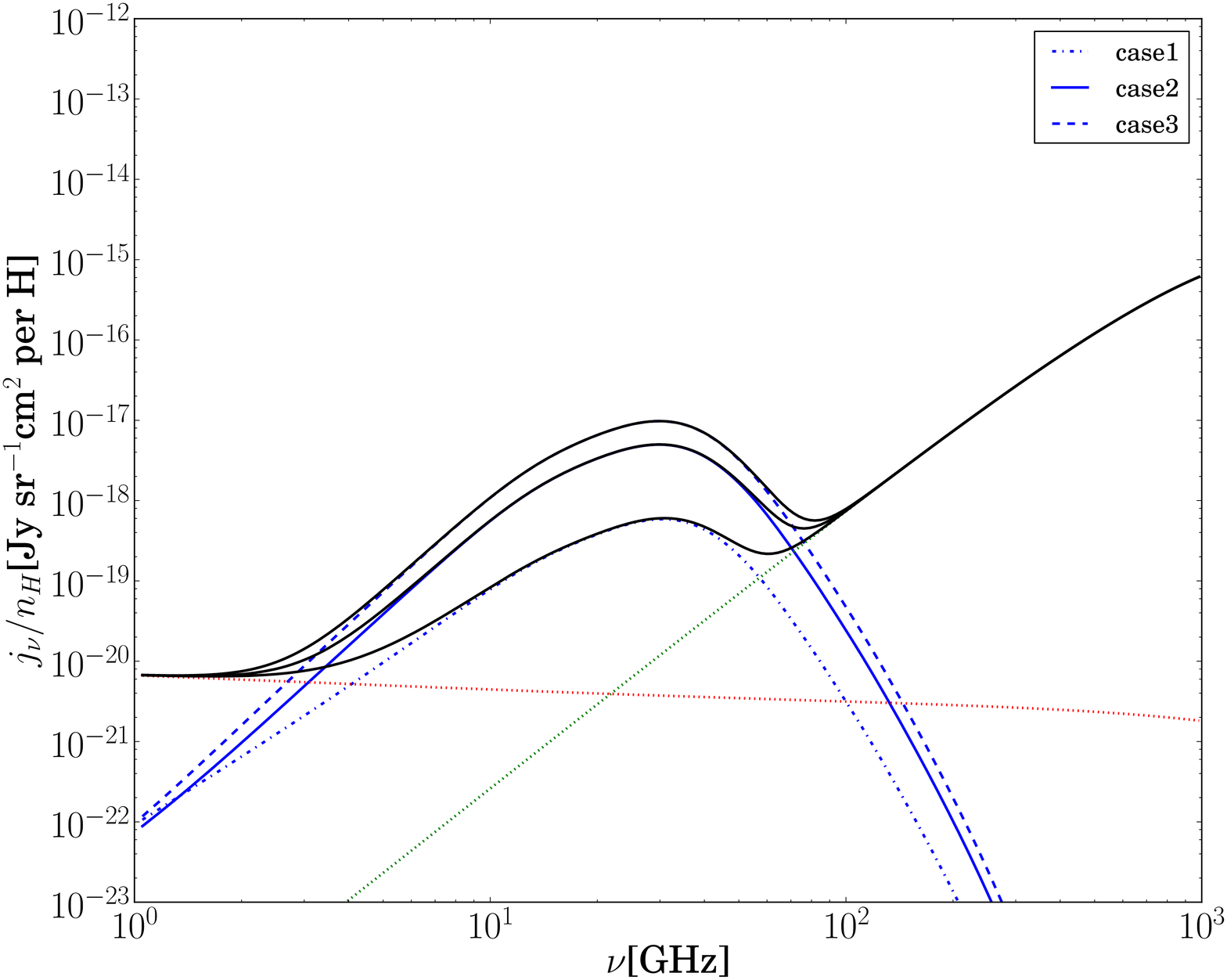}{0.5\textwidth}{(a) CNM}
          \fig{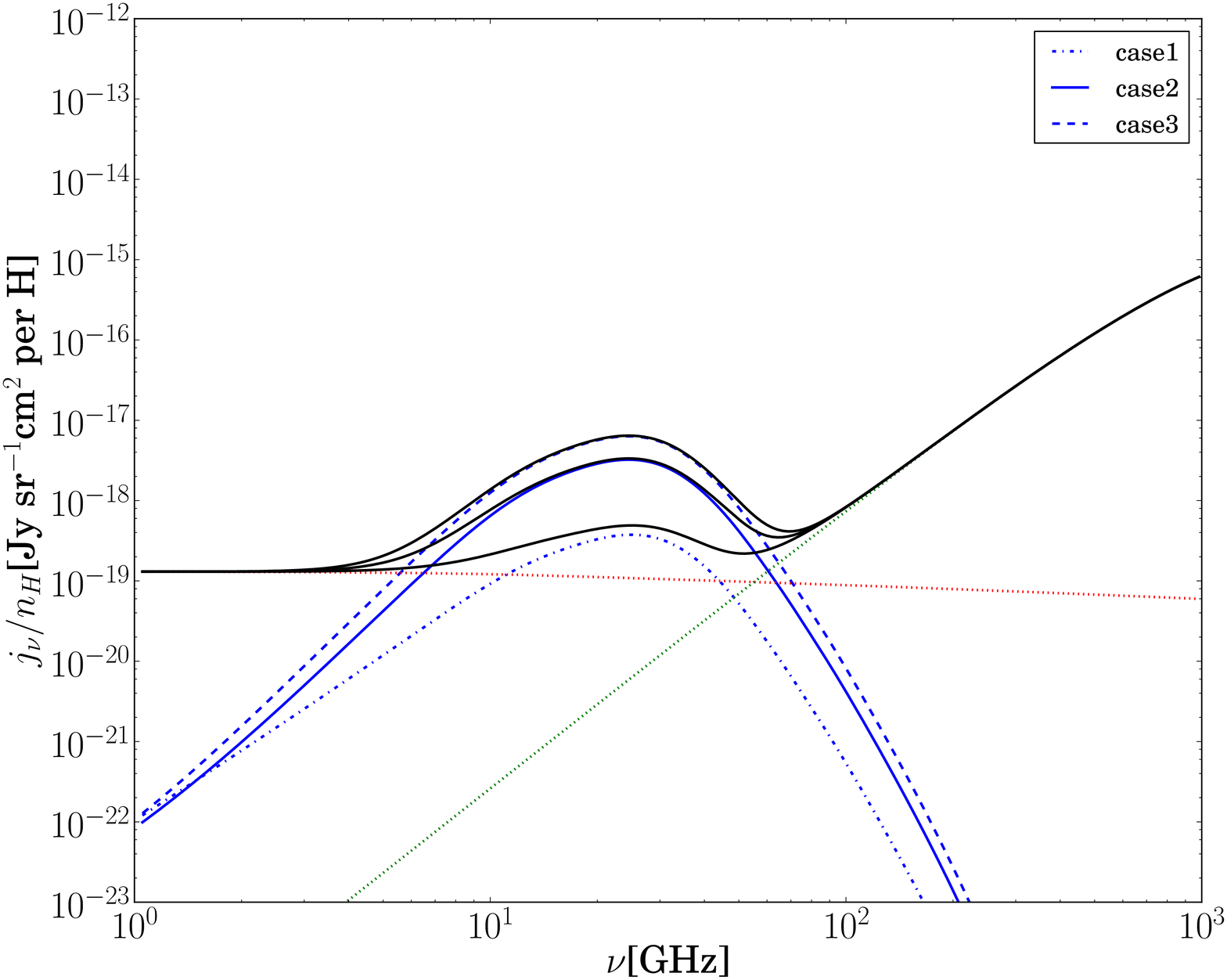}{0.5\textwidth}{(b) WNM}}
\gridline{          
          \fig{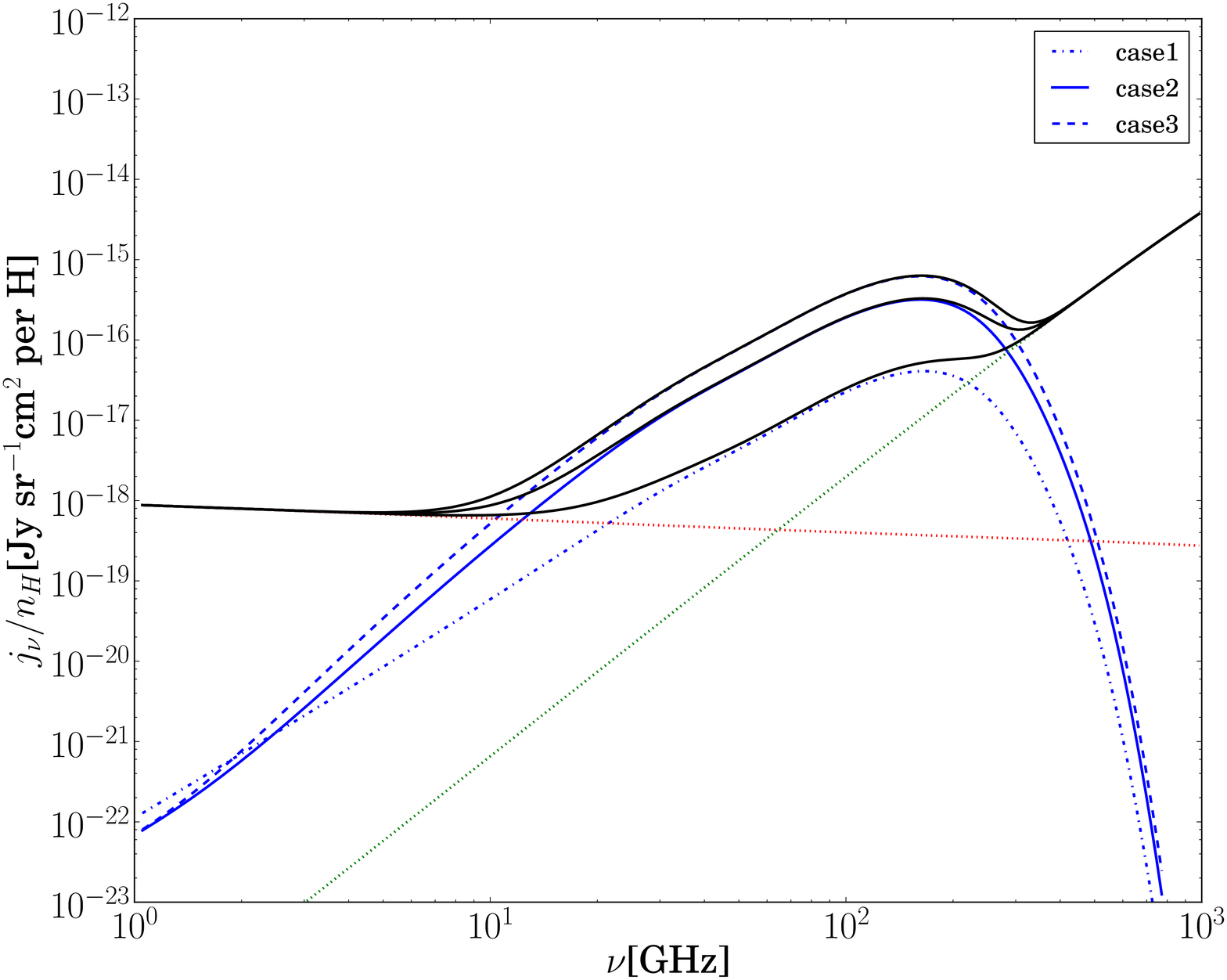}{0.5\textwidth}{(c) PDR}
}
\caption{\textit{Before the impact of the C-shock}: radio-millimeter emissivity including free-free emission, thermal dust emission, and AME for typical ISM conditions for CNM, WNM, and PDR before the impact of the C-shock}
\label{fig:emiss_before}
\end{figure*}
Using \texttt{SpDust}, we compute the AME emissivity \jvsp~(Equation~\ref{eq:emissi}) without including damping and excitation coefficients due to the supersonic neutral drift ($F_{sd}$ and $G_{sd}$ in Equation~\ref{eq:coeffs}). In Figure~\ref{fig:wrot_before}, we show the angular velocity $\omega_{rot}$ (connected blue dots) in Equation~\ref{eq:wrot2} as a function of grain size $a$ based on the computed damping and excitation coefficients for CNM, WNM, and PDR, together with the critical angular velocity $\omega_{cri}$ (red lines) in Equation~\ref{eq:wcri} for a range of maximum tensile strength, $S_{\mbox{\tiny max}}=10^8$ -- $10^{10}$ [erg cm$^{-3}$]. As the grain size becomes smaller, both $\omega_{rot}$ and $\omega_{cri}$ increase. For a grain whose size is smaller than a certain length, where $\omega_{rot}=\omega_{cri}$, one can expect that $\omega_{rot}$ will be larger than $\omega_{cri}$ and grains smaller than that size will be broken apart due to centrifugal force, which, however, is not likely to happen in the cases for the given range of $S_{\mbox{\tiny max}}$ (Figure~\ref{fig:wrot_before}), where the impact of the C-shock is not being considered.

Therefore, the integration over grain size in Equation~\ref{eq:emissi} is performed over the full range (3.5-100\angst). The resulting emissivities, including AME (dotted-dashed, solid and dashed blue line for case1, 2, and 3, respectively), free-free (red dashed line), and thermal dust emission (green dashed line) for CNM, WNM, and PDR, are shown in Figure~\ref{fig:emiss_before}. In addition to the fact that the strength of free-free emissivity increases as we go from CNM to PDR, the AME emissivity is significantly larger than the free-free emissivity for CNM, WNM, and PDR. In particular, for PDR, the peak frequency even moves to the higher frequency ($>100$ GHz) and impacts the emissivity curve at millimeter wavelength that is usually dominated by the emissivity of thermal dust emission if there is no AME.    

\subsection{AME after the Impact of the C-shock}
We compute the AME emissivity \jvsp~(Equation~\ref{eq:emissi}) including damping and excitation coefficients due to the supersonic neutral drift ($F_{sd}$ and $G_{sd}$ in Equation~\ref{eq:coeffs}) by following the formalism described in Section~\ref{sec:dynamics}. The $s_d$ can be varied depending on the exact shock condition in the ISM and we use $s_d=10$ from Figure~\ref{fig:shockprofile}. 

\begin{figure*}
\gridline{\fig{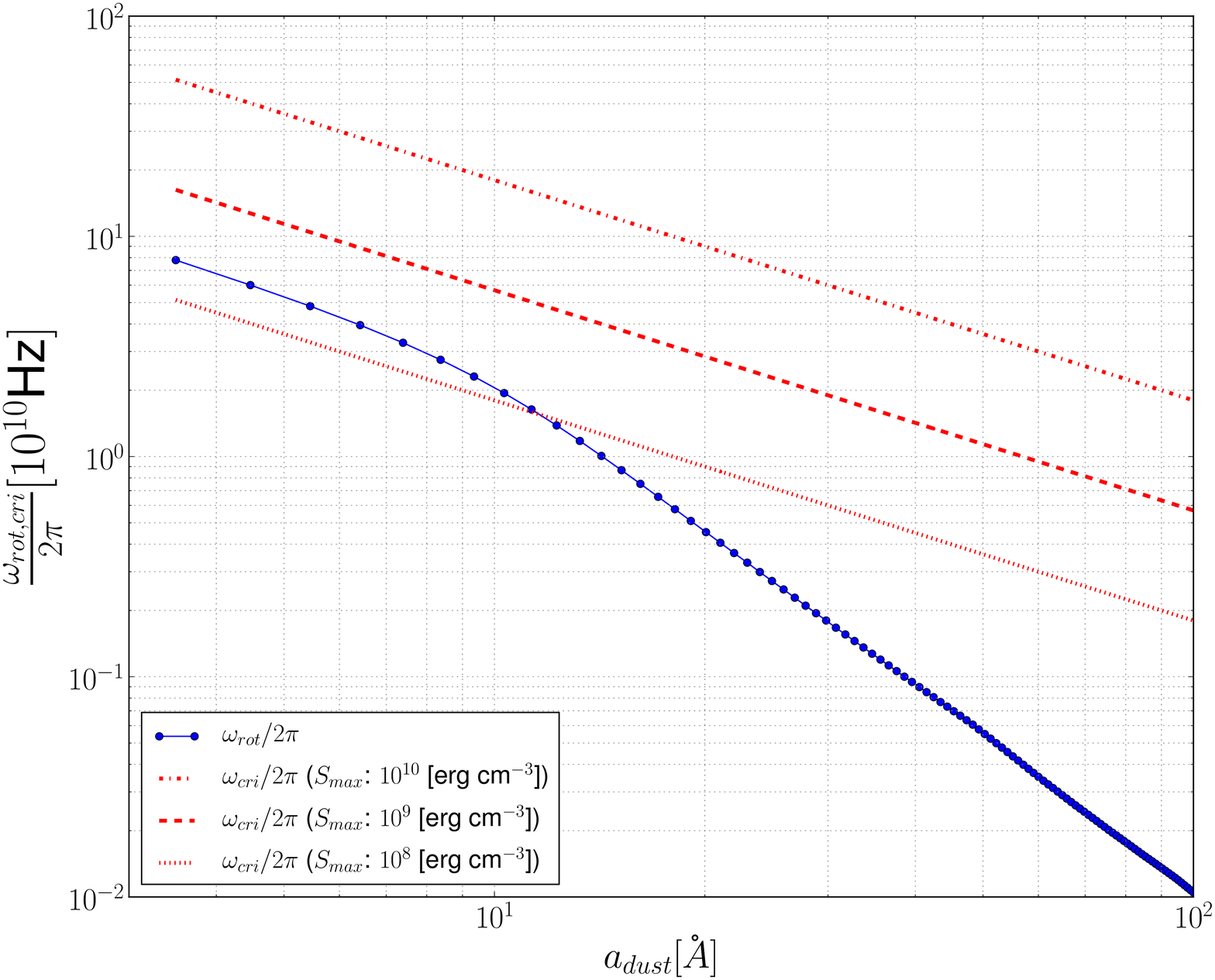}{0.5\textwidth}{(a) CNM}
          \fig{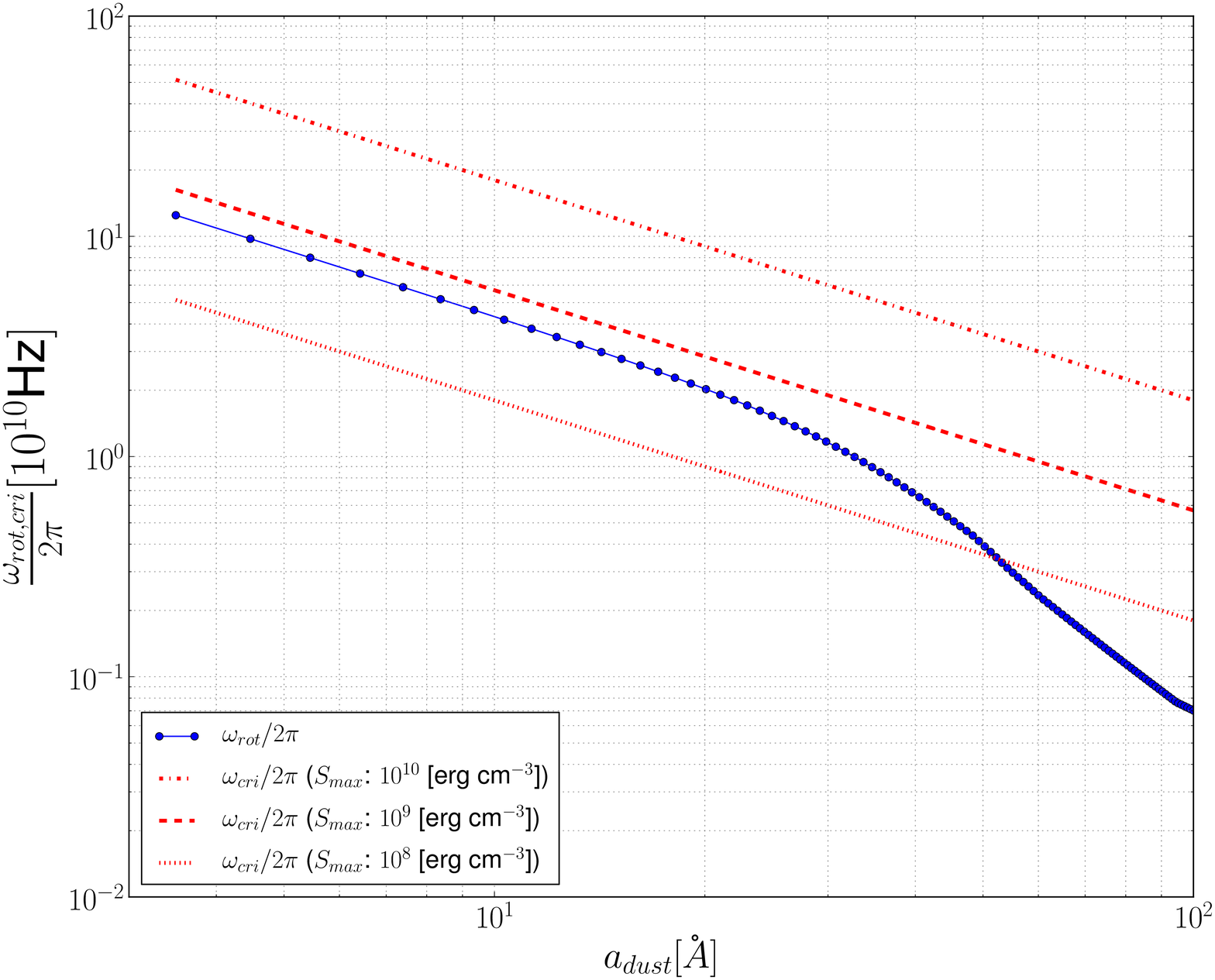}{0.5\textwidth}{(b) WNM}}
\gridline{          
          \fig{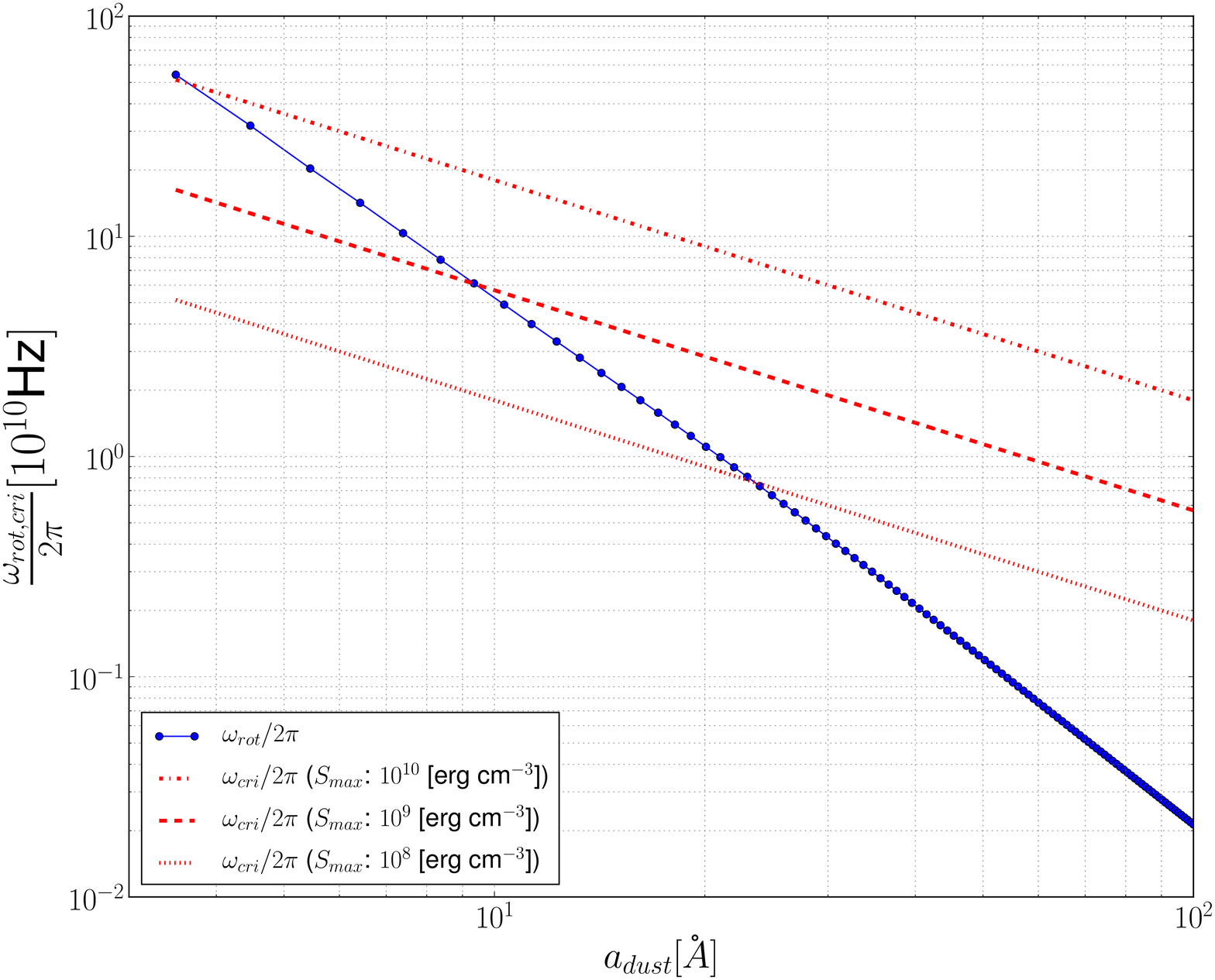}{0.5\textwidth}{(c) PDR}
}
\caption{\textit{After the impact of the C-shock}: angular rotation velocity of dust particles (blue line) emitting AME and critical angular rotation velocity (red lines) for a range of the maximum tensile strength ($10^8$-- $10^{10}$ erg cm$^{-3}$) of the spinning dust particles} 
\label{fig:wrot_after}
\end{figure*}

In Figure~\ref{fig:wrot_after}, we show the angular velocity $\omega_{rot}$ (connected blue dots) as a function of grain size $a$ based on the computed damping and excitation coefficients, including $F_{sd}$ and $G_{sd}$, for CNM, WNM, and PDR, together with the critical angular velocity $\omega_{cri}$ (red lines). Due to the excitation by the interaction with supersonic neutral drift, $\omega_{rot}$ increases significantly, and the critical grain size below which the destruction process happens for the grain with the maximum tensile strength $S_{\mbox{\tiny max}}=10^8$~[erg cm$^{-3}$], becomes larger ($a>1$ nm or $10$\angst). For each ISM condition in Figure~\ref{fig:wrot_after}, we find the critical grain size where $\omega_{rot}$ and $\omega_{cri}$ for $S_{\mbox{\tiny max}}=10^8$~[erg cm$^{-3}$] become equal and use this grain size as \amin\ for integrating Equation~\ref{eq:emissi}, which reduces the AME emissivity.

\begin{figure*}
\gridline{\fig{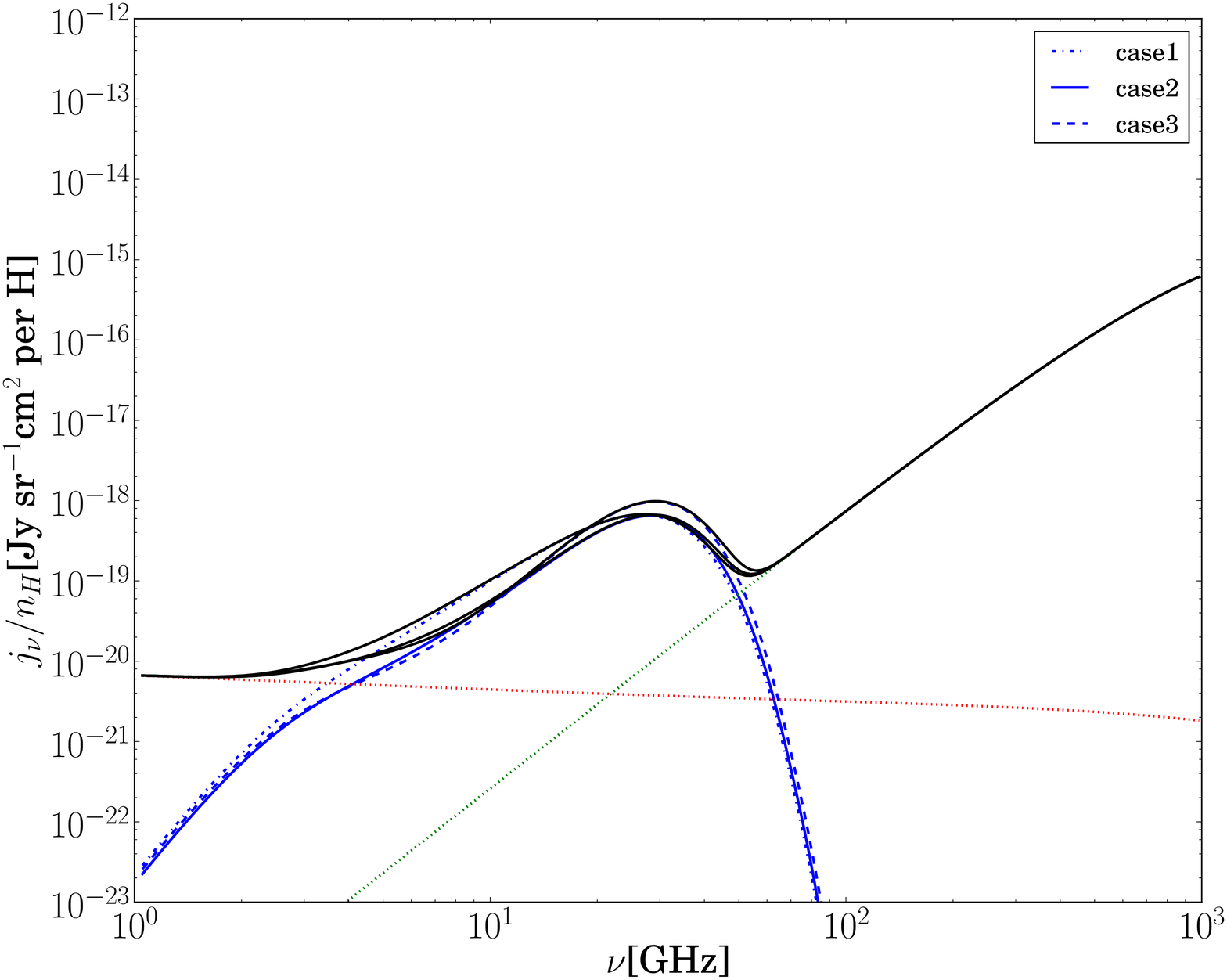}{0.5\textwidth}{(a) CNM}
          \fig{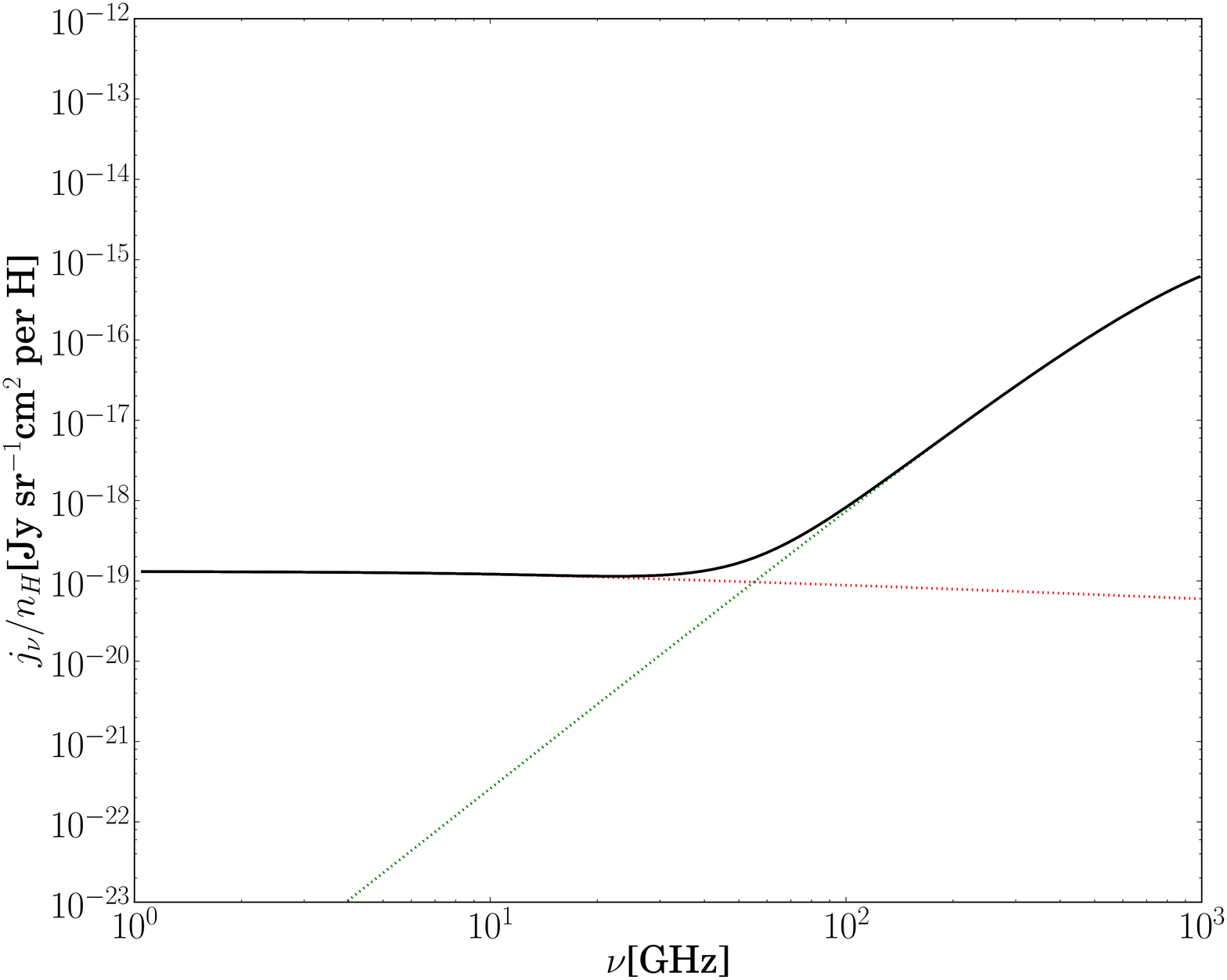}{0.5\textwidth}{(b) WNM}}
\gridline{          
          \fig{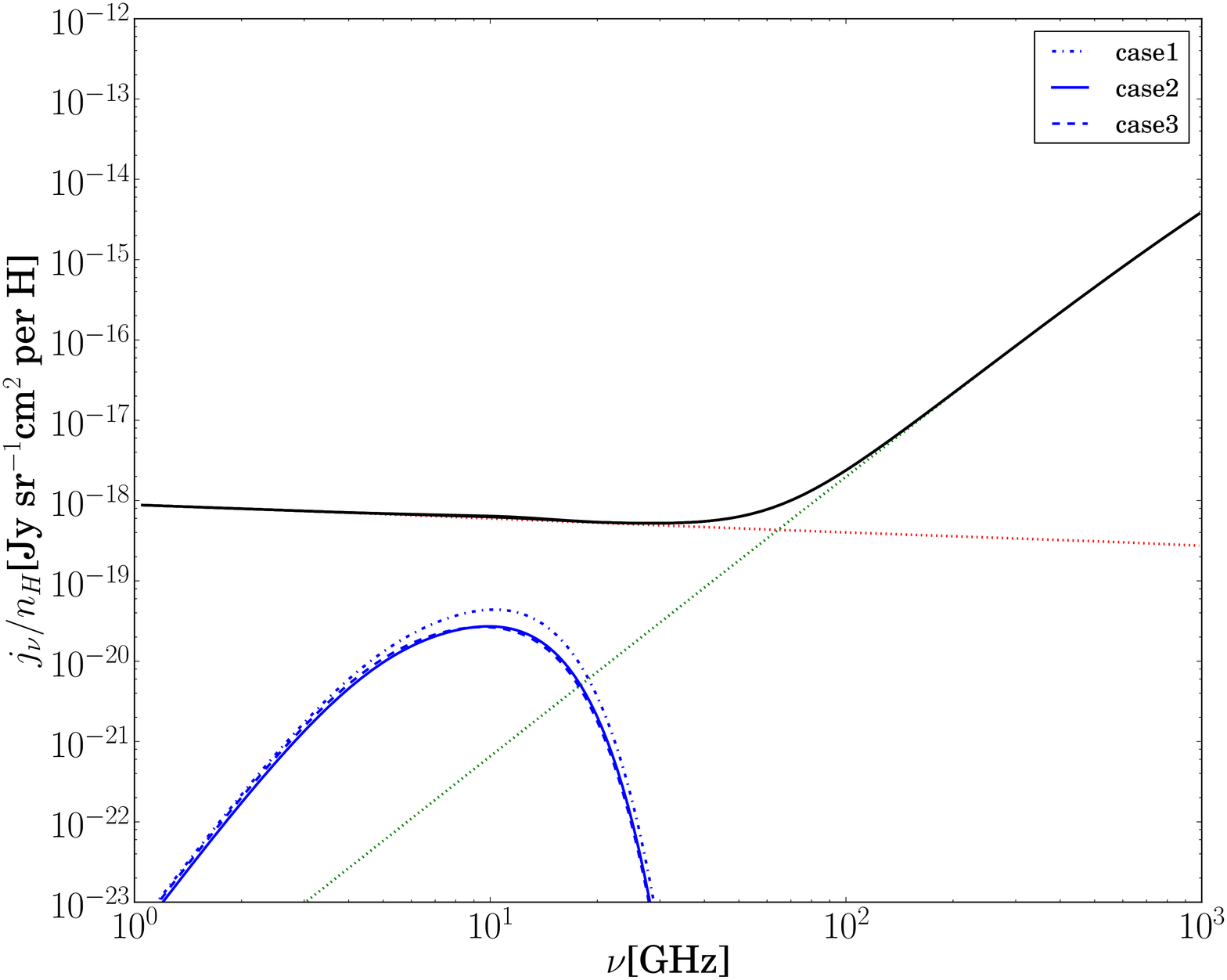}{0.5\textwidth}{(c) PDR}
}
\caption{\textit{After the impact of the C-shock}: radio-millimeter emissivity including free-free emission, thermal dust emission, and AME for typical ISM conditions for CNM, WNM, and PDR after the impact of the C-shock. Note that AME emissivity in the WNM condition is entirely suppressed and negligible compared to free-free and thermal dust continuum emissivity.}
\label{fig:emiss_after}
\end{figure*}

The resulting emissivities, including AME (dotted-dashed, solid, and dashed blue lines for case1, 2, and 3, respectively) impacted by the C-shock, free-free (red dashed line), and thermal dust emission (green dashed line) for CNM, WNM, and PDR, are shown in Figure~\ref{fig:emiss_after}. Compared to Figure~\ref{fig:emiss_before}, we find that the AME emissivity is significantly (for CNM) and almost entirely (for WNM and PDR) suppressed because the small grains are destroyed by centrifugal force from increased spin angular velocity.  


\section{Discussion}\label{sec:discuss}
In Section~\ref{sec:emission}, we demonstrate that the impact of C-shocks can suppress the AME emissivity for typical conditions of the ISM (CNM, WNM, and PDR). Although progress has been made \citep[for a review, see][]{dickinson_etal_2018}, the ISM condition of star-forming regions for strong AME is still not well known; the number of objects that are not detected as AME still have a physical condition similar to those that are detected \citep{scaife_2013}. In this section, we propose a hypothesis of the rise and fall of the AME in star-forming regions according to their evolutionary stage (Seections~\ref{sec:riseame} and \ref{sec:fallame}), consider the `observed' SED predicted from the volume-integrated composite emissivity for monotonically or mildly varying ISM physical parameters (Section~\ref{sec:obsame}), and discuss other rotational disruption processes (Section~\ref{sec:met}), dust fragmentation (Section~\ref{sec:frag}), and the difficulty and prospect of the extragalactic AME detection in the era of high-resolution radio observation (Section~\ref{sec:ngvla}).   
%
%

\subsection{The Rise of AME in the Early Stage of Star Formation}\label{sec:riseame}
A typical lifetime of OB stars is 1-10 Myr. In the early stage of star-formation ($\lesssim 10$ Myr), the strong radiation from young stars exerts a radiative torque on the dust particles and increases their angular velocity \citep{draine_and_weingartner_1997,lazarian_and_hoang_2007}. The radiative torque breaks large grains more efficiently, such that grains with a size larger than $a_{\mbox{\tiny disr}}$ are disrupted by the centrifugal force due to the increased angular velocity \citep{hoang_etal_rat_2019},
\begin{equation}
    \left(\frac{a_{\mbox{\tiny disr}}}{0.1\mu m}\right)^{2.7} \approx 0.046\gamma^{-1} \bar{\lambda}_{0.5}^{-1.7} \left(\frac{U_6}{1.2}\right)^{-1/3} S_{\mbox{\tiny max,9}}^{1/2}
\end{equation} where $\gamma$ is the anisotropy parameter of the radiation field $0<\gamma<1$, and $\bar{\lambda}_{0.5}=(\bar{\lambda}/0.5\mu m)$ for the mean wavelength of the radiation field $\bar{\lambda}$. Here $U_6 = U/10^6$ where $U$ is the radiation energy density normalized by the ISRF energy density at the solar neighborhood ($u_{\mbox{\tiny ISRF}}=5.29\times10^{-14}$ erg cm$^{-3}$ \cite{draine_2011book}). This radiative torque disruption is more efficient than other disruption mechanisms (thermal sublimation, thermal sputtering, grain shattering) in destroying large ($a\gtrsim 0.1\mu m$) grains \citep{hoang_etal_rat_2019}.

Since the radiation field energy density $u_{\mbox{\tiny rad}}$ is $F/c$
where $F$ is flux density $L/(4\pi R^2)$ at a distance $R$ from the source with bolometric luminosity $L$, we can write $U$ as follows:
\begin{equation}\label{eq:radfield}
    U = \frac{u_{\mbox{\tiny rad}}}{u_{\mbox{\tiny ISRF}}} \approx 2100 \left(\frac{L}{10^7 L_\odot}\right)\left(\frac{R}{10\mbox{pc}}\right)^{-2} 
\end{equation}
If a massive star-forming region contains 100 O stars of bolometric luminosity $\sim 10^5 L_{\odot}$ with anisotropic illumination ($\gamma=0.1$) of radiation and $\bar{\lambda}=0.5\mu m$, $a_{\mbox{\tiny disr}}\approx 0.14\mu m$ at 10 pc away and $a_{\mbox{\tiny disr}}\approx 0.25\mu m$ at 100 pc away from the center of the star-forming region for the grain particle with the maximum tensile strength of $S_{\mbox{\tiny max}}=10^9$erg cm$^{-3}$. So, in typical extragalactic star-forming regions with massive star formation, grains with sizes larger than 0.2$\mu m$ (or $2\times 10^{-5}$cm) will be disrupted by the radiation torque, while small nanoparticles ($a\sim10^{-7}$cm) are still resilient against the centrifugal force. 

Detailed dust grain modeling by \cite{silsbee_draine_2016} suggests that `fluffy' grains exposed to solar radiation can be disrupted by increased spin. However, in order to make the radiation pressure an efficient mechanism for breaking large grains, it may require an extreme radiation field (Draine  2021 private communication), which is much larger than the value from Equation~\ref{eq:radfield}. 

The disrupted large grains are likely to be added as small grains and change the shape of the dust size distribution by steepening the power-law distribution in Equation~\ref{eq:asize}. This implies that the AME might be stronger in young (10 Myr or less) and massive ($\sim 100  M_\odot$) star-forming regions before a supernova explosion drives a shock wave into the ISM, which is consistent with the interpretation of the extragalactic AME observation by \cite{murphy_etal_2018}, where the best AME model for the source suggests that AME is from a nascent star-forming region with very young ($<3$ Myr) massive stars.

A compilation of the observations of AME from Galactic star-forming regions associated with \HII\ region is reported in \cite{dickinson_2013}. It is difficult to compare the reported values of the significance of AME detection for individual regions because the observations are made with different angular scales and the exact definitions of the significance of AME detection are different. However, amongst the observed star-forming regions, those that show a significant flux enhancement over the thermal free-free emission in $\sim 30$ GHz frequency are powered by OB stars \citep{finkbeiner_etal_2004,dickinson_etal_2009,planck_2011,shuping_etal_2012,tibbs_etal_2012}. On the other hand, the AME is not detected from a star-forming region that is relatively older and previously impacted by shocks, like the Orion Nebula \citep{planck_2011}, and from the supernova remnant like 3C 396 \citep{cruciani_etal_2016}. These observations suggest that another important environmental issue for understanding the observed AME might be whether or not the ISM is impacted by a shock, which is correlated with the age of the star-forming region.

\subsection{The Fall of AME in the Later Stage of Star Formation}\label{sec:fallame}
As we show in Section~\ref{sec:emission}, the C-shock in the magnetized medium can suppress the AME by disrupting small nanoparticles ($a \sim 10^{-7}$ cm). This implies that if a supernova explodes within 10-100 Myrs when massive stars become old enough to die, the star-forming region will be impacted by the C-shock and may not be the source of strong AME anymore. Here we argue that this scenario is plausible by comparing several important timescales: shock propagation time, neutral flow time, rotational disruption time, and dust reformation time. 

Shock propagation time is the time for the shock front with speed $v_{s}$ to propagate into the ISM to a distance $l$ and can be written as 
\begin{equation}
    \tau_{\mbox{\tiny prop}} = 4.8\times10^5\left(\frac{l}{10\mbox{pc}}\right)\left(\frac{20\mbox{km/s}}{v_s}\right) \mbox{yr}.
\end{equation}
Neutral flow time is the time for neutrals with drift velocity $v_{\mbox{\tiny drift}}$ to pass the length scale of shock $l_s$ and can be written as \citep{hoang_etal_2019}
\begin{equation}
    \tau_{\mbox{\tiny flow}}=30\left(\frac{l_s}{10^{15}\mbox{cm}}\right)\left(\frac{10\mbox{km/s}}{v_{\mbox{\tiny drift}}}\right) \mbox{yr}.
\end{equation}
Rotational disruption time is the time required to spin-up nanoparticles with maximum tensile strength $S_{\mbox{\tiny max}}$ to $\omega_{cri}$ and can be written as \citep{hoang_etal_2019}
\begin{equation}
    \tau_{\mbox{\tiny disr}}\simeq0.005a_{-7}^4\left(\frac{\nh}{10^5\mbox{cm}^{-3}}\right)^{-1}S_{\mbox{\tiny max,9}}\left(\frac{v_{\mbox{\tiny drift}}}{10\mbox{km/s}}\right)^{-3} \mbox{yr}.
\end{equation}
Dust reformation time is the time required to replenish the disrupted nanoparticles, which is most likely done by the fresh dust formation process via stellar winds from the low- and intermediate-mass stars evolved from the main sequence because dust growth via coagulation is a very slow process with gigayears of timescale \citep{asano_etal_2013}. In particular, the thermally pulsating asymptotic giant branch (AGB) phase of intermediate-mass stars is likely to be the most promising site for dust formation \citep{morgan_and_edmunds_2003}. Given that the main-sequence lifetime of intermediate-mass ($M \sim 3-5M_{\odot}$) stars is $\gtrsim 100$ Myr \citep{schaller_etal_1992}, the dust reformation time via stellar wind from evolved stars is at least larger than 100 Myr, $\tau_{\mbox{\tiny reform}}\gtrsim100~\mbox{Myr}$, and this will probably be the fastest way of replenishing the nanoparticles.

To understand the effect of gas bombardment on disruption of small nanoparticles in C-shocks, we need to compare $\tau_{\mbox{\tiny disr}}$ with $\tau_{\mbox{\tiny flow}}$:
\begin{equation}
\begin{aligned}
    \frac{\tau_{\mbox{\tiny flow}}}{\tau_{\mbox{\tiny disr}}}\simeq6000a^{-4}_{-7}\left(\frac{l_s}{10^{15}\mbox{cm}}\right) \left(\frac{\nh}{10^5\mbox{cm}^{-3}}\right)S_{\mbox{\tiny max,9}}^{-1}\\\times\left(\frac{v_{\mbox{\tiny drift}}}{10\mbox{km/s}}\right)^{2}.
\end{aligned}
\end{equation}
In the shock layer ($l_s\approx 10^{15}$cm), the $\frac{\tau_{\mbox{\tiny flow}}}{\tau_{\mbox{\tiny disr}}}$ for CNM and PDR is significantly larger than 1 for typical values of nanoparticle size $a \approx 10^{-7}$cm, drift velocity $v_{\mbox{\tiny drift}}\approx 10$km/s and maximum tensile strength $S_{\mbox{\tiny max}} \approx 10^9$ erg cm$^{-3}$. Although, for WNM, $\frac{\tau_{\mbox{\tiny flow}}}{\tau_{\mbox{\tiny disr}}}$ is less than 1 ($\frac{\tau_{\mbox{\tiny flow}}}{\tau_{\mbox{\tiny disr}}}\sim0.024$), for the same values of $a$, $v_{\mbox{\tiny drift}}$ and $S_{\mbox{\tiny max}}$, $\tau_{\mbox{\tiny flow}}$ and $\tau_{\mbox{\tiny disr}}$ become easily comparable as the particle size becomes smaller ($a_{-7}<1$) and $v_{\mbox{\tiny drift}}$ becomes larger (due to a  high-velocity shock). Therefore, a $\tau_{\mbox{\tiny flow}}$ much larger than $\tau_{\mbox{\tiny disr}}$ implies that the grain disruption process is almost instantaneous after the shock sweeps the ISM. 
 
Even though the nanoparticles are quickly disrupted in the shock layer ($\frac{\tau_{\mbox{\tiny flow}}}{\tau_{\mbox{\tiny disr}}}\gg 1$), AME will not be suppressed if the small nanoparticles, after the shock propagation, are reformed quickly as a result of the dust formation process via stellar wind from AGB stars. However, the shock propagation time scale, $\tau_{\mbox{\tiny prop}}$ is much shorter than the dust reformation time $\tau_{\mbox{\tiny reform}}\gtrsim100~\mbox{Myr}$. It implies that the shock propagates almost instantaneously through the ISM and disrupts the nanoparticles without quick replenishment of dust grains by the AGB star.

Dust grains are also produced in the supernova explosion \citep[for a review, see][]{sarangi_etal_2018}. But in such high-temperature gas ($T>10^6$K), a 0.1$\mu$m size grain survives for $<0.1$Myr \citep{draine_2011book}, and the theoretical modeling works of dust formation and evolution in the supernova explosion suggest that the dust produced in the supernova explosion is mostly large grains \citep[$>1\mu$m;][]{brooker_etal_2021,sarangi_etal_2018}. Also, \cite{gall_etal_2014} reported that only grain size distributions with grain radii larger than 0.25$\mu$m with a lower limit of 0.7$\mu$m can reproduce the observed supernova extinction curves. The large dust grains formed in the high-temperature supernova explosion do not survive long and are not likely to be the source of AME.

\subsection{Observed Radio-millimeter SED}\label{sec:obsame}
What we compute is the emissivity, not the observed flux density, as we note in Section~\ref{sec:emission}. However, we can infer the significance of the observed AME relative to the free-free and thermal dust emission by assuming a simple profile of the physical parameters. 

In a simple spherical geometry, the observed SED of the star-forming region at distance $D$ is determined by the measured flux density $F_\nu$ at $\nu$.
\begin{equation}\label{eq:obsemiss}
    F_\nu = \frac{1}{4\pi D^2}\int^{R_f}_{R_i}dr4\pi r^2\nh \left(\frac{4\pi j_\nu}{\nh}\right)
\end{equation}
where 
\begin{equation}
\frac{4\pi j_\nu}{\nh} = 
    \left(\frac{4\pi \jvff}{\nh} + \frac{4\pi \jvbb}{\nh} + \frac{4\pi \jvsp}{\nh} \right)
\end{equation}
and the volume integration is performed from the inner radius $R_i$ to the outer radius $R_f$ of the star-forming region.

From Equation~\ref{eq:obsemiss} and the emissivity equation for each emission (Equation~\ref{eq:jvff}, \ref{eq:jvbb} and \ref{eq:emissi}), we can infer that the observed flux density of free-free emission depends on the radial distribution of \nh\ and $T$, the flux density of thermal dust emission depends on the radial distribution of \nh\, and the flux density of AME depends on the radial distribution of \nh, $T$, $\chi$, $x_{\mbox{\tiny H}}$.

In general, for each ISM, the CNM, WNM, and PDR around the star-forming regions, $\nh(r)$, $T(r)$, $\chi(r)$, and $x_{\mbox{\tiny H}}(r)$, are monotonically decreasing and not likely to vary much. Therefore, the `relative' strength of the emissivity for free-free and thermal dust emission and AME will not change much with $r$. For each ISM condition (CNM, WNM, and PDR), we indeed made $\nh(r)$, $T(r)$, $\chi(r)$, and $x_{\mbox{\tiny H}}(r)$ an order of magnitude lower than the typical values and computed the total emissivity. We find that although the peak of AME emissivity changes, the relative strength of each emission does not change much by the variation of these parameters. 

This implies that the observed radio-millimeter SED resulting from the volume integration of the emissivity reflects the relative strength of the emissivity for each emission, and the relative strength of each emission in the radio-millimeter SED is not much altered by the spatial variation of the ISM physical parameters. This justifies our argument; for the smooth and monotonic variation of ISM parameters, the shape of the observed radio-millimeter SED, a result of volume integration of the free-free, thermal dust, and AME emissivity, is determined by the relative contribution of each emissivity to the total emissivity.

\subsection{Regular Mechanical Torque by Subsonic Drift}\label{sec:met}
In this study, we assume spherical dust affected by a stochastic mechanical torque from supersonic drift of dust relative to the neutral atoms \citep{hoang_etal_2019}. However, realistic nanoparticles are expected to be irregular and have helicity, which increases the spin angular momentum by the interaction with subsonic astrophysical flow \citep{lazarian_and_hoang_2007a,hoang_etal_2018}. This mechanical torque (called MET) was originally proposed for grain alignment \citep{lazarian_and_hoang_2007a} and known to be much more efficient for spinning up nanoparticles than stochastic mechanical torque \citep{lazarian_and_hoang_2021}. Our understanding of the impact of MET is primarily based on qualitative guidance from a simple model \citep{lazarian_and_hoang_2007a}; however, more complicated models and analyses \citep[e.g.,][]{das_and_weingartner_2016,hoang_etal_2018,reissl_etal_2022} support the idea that irregular grains exhibit helicity while interacting with gaseous flows, and the corresponding regular torques dominate the stochastic mechanical torques \citep{lazarian_and_hoang_2021}. The MET is difficult to calculate and depends strongly on grain shape \citep{hoang_lazarian_2018}, and there is no available tool for numerical studies to characterize its performance in astrophysical situations \citep{lazarian_and_hoang_2021}. Although it is hard to make a quantitative argument on the impact of MET on dust disruption, the efficiency of rotational disruption by the centrifugal force due to increased spin from mechanical torque (either the regular or the stochastic mechanical torque, or both) may be perhaps more efficient in real astronomical situations \citep{hoang_etal_2019}.   

\subsection{Grain Fragmentation}\label{sec:frag}
In this study, we assume that small nanoparticles with higher angular velocity than the critical angular velocity for centrifugal disruption are instantaneously and completely destroyed. However, because the rotational disruption process cannot destroy grains to atoms, one can expect a continuous hierarchical disruption where large grains become small grains, and as a result, the grain size distribution might change \citep[e.g.,][]{guillet_etal_2011}. Our study does not model the fragmentation of grains. If the grains are disrupted to smaller grains and those small grains survive for long enough time, one can expect increased AME. However, we argue that smaller grains fragmented from large grains with an angular momentum larger than the critical value have higher angular velocity and will be disrupted quickly.

Let us suppose that one large spherical grain with angular velocity $\omega$ and moment of inertia $I=\frac{2}{5}MR^2$ for mass $M$ and radius $R$ has angular momentum $J=I\omega$ and fragments into two equal-sized spherical grains. Then, the two small grains will have the same angular momentum, 
$J^{\prime}=\frac{2}{5}\left(\frac{M}{2}\right){R^\prime}^2{\omega^\prime}$. If one assumes that the angular momentum is conserved ($J=2J^\prime$), $\omega^\prime = \omega \left(\frac{R}{R^\prime}\right)^2$. In this simple fragmentation scenario, $\omega$ is inversely proportional to the square of the dust size ($\omega\sim\frac{1}{a^2}$), which increases faster than the critical angular velocity $\omega_{cri}$ in Equation~\ref{eq:wcri} as the dust size decreases. Although this simple fragmentation scenario does not represent reality, it is likely that small grains fragmented from large grains have a higher angular velocity than $\omega_{cri}$ and therefore are disrupted easily.

\subsection{AME from Extragalactic Star-forming Region in the Era of ngVLA}\label{sec:ngvla}
Detection of AME from extragalactic star-forming regions is rare (reported for only two galaxies). Assuming that there is no difference in the ISM condition, on average, in our Milky Way and other galaxies, the following observational selection effects affecting the detection of \HII\ regions and AME may explain the rarity of extragalactic AME.

(i) The AME detection experiment for other galaxies starts from the existing radio map at the frequency where thermal free-free emission is dominant \citep[e.g.,][]{eric_sfr_2018}. For the radio sources detected in the map, a follow-up multifrequency observation is made if the radio spectral SED of the source is unusually shallow or suggests an increased flux at higher frequency. It means that detecting the star-forming region in radio should not be biased to an extreme system. However, detecting thermal free-free emission from extragalactic \HII\ regions has systematic selection effects: beam dilution, confusion with disk emission, and confusion with a nonthermal discrete source \citep{israel_1980}. These systematic effects make the detection of \HII\ regions difficult because the contrast between the individual \HII\ region and the underlying extended disk emission of the galaxy becomes small. As a result, the detected extragalactic \HII\ regions with radio free-free emission may be biased towards giant \HII\ regions formed by massive stars, as illustrated by \cite{israel_etal_1975}, showing that at better resolution, the detected \HII\ regions at larger beams break up into groups or chains of smaller clumps. Therefore, the number of detected extragalactic \HII\ regions with free-free emission would be less than that of our Milky Way, where we see many individual (small and large) \HII\ regions, which limits our ability to find AME candidates. 

(ii) The observations suggest that AME is likely to be a local phenomenon. For example, recent studies of the spatially resolved AME region show that the 31GHz emission (as a proxy for AME) is related to the local PAH column density \citep{arcetord_etal_2020}, and the ``excess'' emission at 15 GHz (interpreted as AME) larger than the ``expected'' radio synchrotron and free-free emission varies locally \citep{battistelli_etal_2015}. If the AME is governed by local ISM conditions on small scales, one expects to see a smaller number of AME sources in other galaxies than our Milky Way because the same radio beam encompasses a much larger area ($\gtrsim 100$pc) in nearby galaxies, and the emission in the beam including \HII\ regions is dominated by free-free emission from the \HII\ regions, not by AME from the surrounding molecular clouds.      
In addition to the dust destruction mechanism proposed in this study, the systematic effect in radio observations makes the detection of extragalactic AME difficult. Therefore, if detected, the extragalactic AME is likely to be from a large molecular cloud around a young massive star cluster before a supernova explodes ($<10$Myr), which also explains why extragalactic AME is not detected very often compared to our Milky Way, where AME is detected more frequently.

When the ngVLA operates with high angular resolution and sensitivity \citep[][]{selina_etal_2018}, we will be able to locate individually resolved extragalactic star-forming regions (at a similar scale as our Milky Way) and find AME candidates. We can follow up on them with multifrequency observations to confirm the SED shape of AME, which will enable us to test the AME formation hypothesis based on large samples.

\section{Summary}\label{sec:summary}
We consider the impact of C-shocks on the rotational grain destruction process for small nanoparticles emitting AME \citep{hoang_etal_2019} and compute the emissivity of AME, free-free, and thermal dust continuum emission in the radio-millimeter wavelength for the typical CNM, WNM, and PDR ISM conditions surrounding star-forming regions. Our model of the rotational destruction process for AME suppression is based on a spherical dust shape and a specific range of the maximum tensile strength to withstand centrifugal stress, which is very poorly understood. With these caveats in mind, our study can be summarized as follows.
\begin{enumerate}
    \item[$\bullet$] The magnetized shock (C-shock) for the typical conditions of CNM, WNM, and PDR can create a supersonic neutral drift that collides with grains and increases their spin angular momentum.
    \item[$\bullet$] If the spin angular velocity is larger than the critical angular velocity that is resilient to the centrifugal force for given maximum tensile strength, the AME emitting grains break up, and the emissivity of AME is suppressed relative to the others: emissivity of free-free and thermal dust continuum emission.
    \item[$\bullet$] If a C-shock destroys the small nanoparticles in the ISM surrounding the star-forming region, the AME that might be prominent in the early stage of star-formation ($\lesssim 10$Myr) becomes significantly suppressed and might not be detectable after $\sim 10$ Myr, when the shock from the supernova explosion develops and impacts the surrounding ISM in a massive star-forming region, which might explain the rare observations of extragalactic AME.
    \item[$\bullet$] This study suggests that the presence of AME can be indicative of an early stage of star formation, and the strength of AME depends on the local conditions of the ISM, which implies that spatially resolved high-resolution radio observations are required to understand the detailed physics of AME and its connection to the ISM.
\end{enumerate}

\acknowledgments
I.Y. is grateful to the referee who provided valuable comments that greatly improved the current work.
I.Y. thanks Bruce Draine for his valuable comment on the \HII\ region and the grain destruction process and Antoine Gusdorf and Sylvie Cabrit for answering questions regarding the shock code. I.Y. also thanks Eric Murphy and Sean Linden for useful conversations regarding the observations of extragalactic AME. I.Y. acknowledges kind financial support from the National Radio Astronomy Observatory for the publication of this work. The National Radio Astronomy Observatory is a facility of the National Science Foundation operated under cooperative agreement by Associated Universities, Inc.

\software
{\texttt{SpDust} \citep{spdust2009,silsbee_etal_2011}, \texttt{Paris-Durham} shock code \citep{flower_etal_2003,lesaffre_etal_2013,godard_etal_2019}, \texttt{numpy} \citep{harris_etal_2020}, \texttt{matplotlib} \citep{hunter_2007}}

\vspace{1mm}
\bibliography{ame}{}
\bibliographystyle{aasjournal}



\end{document}